\documentclass[]{aastex631}

\usepackage{natbib}
\bibpunct{(}{)}{;}{a}{}{,} 

\usepackage{graphicx}
\usepackage{bm}

\usepackage{amsmath}
\usepackage{tabularx}
\usepackage{multirow}

\usepackage{savesym}
\savesymbol{tablenum}
\usepackage{siunitx}
\restoresymbol{SIX}{tablenum}
\usepackage{placeins}
\usepackage{hyperref}
\usepackage{ulem}

\DeclareSIUnit[]\rsun
{\text{R\ensuremath{_{\textup{sun}}}}}

\newcommand{\hrieuv}{HRI\textsubscript{EUV}}
\newcommand{\hrilya}{HRI\textsubscript{Lya}}

\DeclareMathOperator{\sinc}{sinc}
\usepackage{booktabs} 

\begin{document} 

    \title{A Point-Spread Function for the Extreme Ultraviolet High-Resolution Imager on board Solar Orbiter}
    \author[0000-0001-7662-1960]{Stefan J. Hofmeister}
    \affiliation{Columbia Astrophysics Laboratory, Columbia University, 538 West 120th Street, New York, NY 10027}
  
    \author[0000-0000-0000-0000]{Emil Kraaikamp}
    \affiliation{Solar-Terrestrial Centre of Excellence – SIDC, Royal Observatory of Belgium, Ringlaan -3- Av. Circulaire, 1180 Brussels, Belgium}

    \author[0000-0000-0000-0000]{Sergei Shestov}
    \affiliation{Solar-Terrestrial Centre of Excellence – SIDC, Royal Observatory of Belgium, Ringlaan -3- Av. Circulaire, 1180 Brussels, Belgium}
    \affiliation{Centre Spatial de Li\`ege, Universit\'e de Li\`ege, Av. du Pr\'e-Aily B29, 4031 Angleur, Belgium}

    \author[0000-0001-7298-2320]{Luca Teriaca}
    \affiliation{Max Planck Institute for Solar System Research, Justus-von-Liebig-Weg 3, 37077 G\"{o}ttingen, Germany}

    \author[0000-0001-7662-1960]{Alexandros Koukras}
    \affiliation{Columbia Astrophysics Laboratory, Columbia University, 538 West 120th Street, New York, NY 10027}

    \author[0000-0000-0000-0000]{Cis Verbeeck}
    \affiliation{Solar-Terrestrial Centre of Excellence – SIDC, Royal Observatory of Belgium, Ringlaan -3- Av. Circulaire, 1180 Brussels, Belgium}

    \author[0000-0000-0000-0000]{Frederic Auchere}
    \affiliation{Université Paris-Saclay, CNRS, Institut d'Astrophysique Spatiale, 91405, Orsay, France}

    \author[0000-0002-1111-6610]{Daniel W. Savin}
    \affiliation{Columbia Astrophysics Laboratory, Columbia University, 538 West 120th Street, New York, NY 10027}
   
    \author[0000-0001-7748-4179]{Michael Hahn}
    \affiliation{Columbia Astrophysics Laboratory, Columbia University, 538 West 120th Street, New York, NY 10027}

    \author[0000-0000-0000-0000]{David Berghmans}
    \affiliation{Solar-Terrestrial Centre of Excellence – SIDC, Royal Observatory of Belgium, Ringlaan -3- Av. Circulaire, 1180 Brussels, Belgium}

   \date{\today}

  \begin{abstract}
We present the point-spread function (PSF) of the Extreme Ultraviolet High-Resolution Imager (\hrieuv) onboard Solar Orbiter, which observes the Sun at \SI{174}{\angstrom}. This PSF provides a quantitative description of light diffracted by the mesh and mounting supporting the entrance filter, light diffracted by the mesh supporting the filter-wheel filter, as well as light that is diffusely scattered by the micro-roughness of the mirrors. Deconvolution with this PSF corrects the images for instrumental scattered light, substantially improving image quality and photometric accuracy.

First, we determine the diffraction component of the PSF from mechanical drawings of the instrument. We find that \SI{26}{\percent} of the incoming light is diffracted, predominantly by the entrance-filter mounting and mesh. Second, we fit the diffuse scattered light using partial image occultations during the 2023-Jan-03 Mercury transit. We find that the diffuse scattered light is well described by a softened power law, which scatters \SI{42}{\percent} of light over the detector. Combined, \SI{57}{\percent} of the incoming light is redistributed over the detector by diffraction and scattering.

Correcting for these effects markedly enhances the dynamic range and contrast of the observations. The intensity in bright structures intensifies by up to \SI{40}{\percent} and the intensity in dark structures decreases by up to \SI{85}{\percent}. All images features become much clearer, facilitating a more precise scientific analysis of \hrieuv\ observations.

   \end{abstract}


\section{Introduction}

In this work, we derive the point-spread function (PSF) of the high-resolution EUV imager \hrieuv, one of the instruments of the Extreme Ultraviolet Imager (EUI) on board Solar Orbiter. EUI consists of two high-resolution imagers (HRIs) and a dual-channel full-Sun imager (FSI), with \hrieuv\  operating at \SI{174}{\angstrom}, \hrilya\  at \SI{1216}{\angstrom}, and FSI at \SI{174} and \SI{304}{\angstrom} \citep{rochus2020}. At these wavelengths, instruments typically scatter a significant portion of the incoming light across the detector, degrading image resolution and dynamic contrast. 

The instrumental PSF describes the redistribution of light caused by the instrument design and mechanical imperfections. Light scattered over small angles primarily results in image blurring, while larger-angle scattering reduces dynamic contrast. By deconvolving the collected images with the instrumental PSF, one can reconstruct a more accurate representation of the true scene and mitigate these instrumental effects.

Here, we derive the instrumental PSF for \hrieuv\ using Mercury transit observations, enabling an accurate correction of its images. Since \hrilya\ is currently affected by instrumental limitations and FSI does not provide sufficient spatial resolution, the same analysis cannot be applied to the other EUI channels.

The PSF of \hrieuv\ primarily consists of two components: (1) a complex diffraction pattern caused by the mountings and meshes supporting the filters at the entrance and filter wheel, resulting in directional diffraction; and (2) an isotropic, diffuse component that redistributes light across the detector, likely caused by microroughness of the mirrors. We characterize the PSF of \hrieuv\ by calculating the diffraction from the mechanical drawings and calibrate the diffuse scattered light using partial solar occultation images from the 2023~Jan~03 Mercury transit, following the methodology described in \citet{hofmeister2022} and \citet{Hofmeister2024_psfs}. We then  evaluate the accuracy of the derived PSF and analyze its effect on collected images.

The rest of the study is structured as follows. Section~\ref{sec:instrument} describes the \hrieuv\ instrument. Section~\ref{sec:observations} presents the Mercury transit observations. Section~\ref{sec:first_guess} derives a first guess for the expected amount of scattered and diffracted light using the Mercury transit images. Section~\ref{sec:diffraction} derives the diffraction pattern from the mechanical drawings. Section~\ref{sec:diffuse} calibrates the amount of diffuse scattered light from the Mercury transit images. Section~\ref{sec:finalPSF} presents the final PSF, provides guidance on its application to \hrieuv\ images, and evaluates the quality of this PSF. Section~\ref{sec:discussion} discusses our results with respect to the instrument design. Section~\ref{sec:conclusions} summarizes our study.

\section{The instrument} \label{sec:instrument}

\begin{figure}
    \centering
    \includegraphics[width=\textwidth]{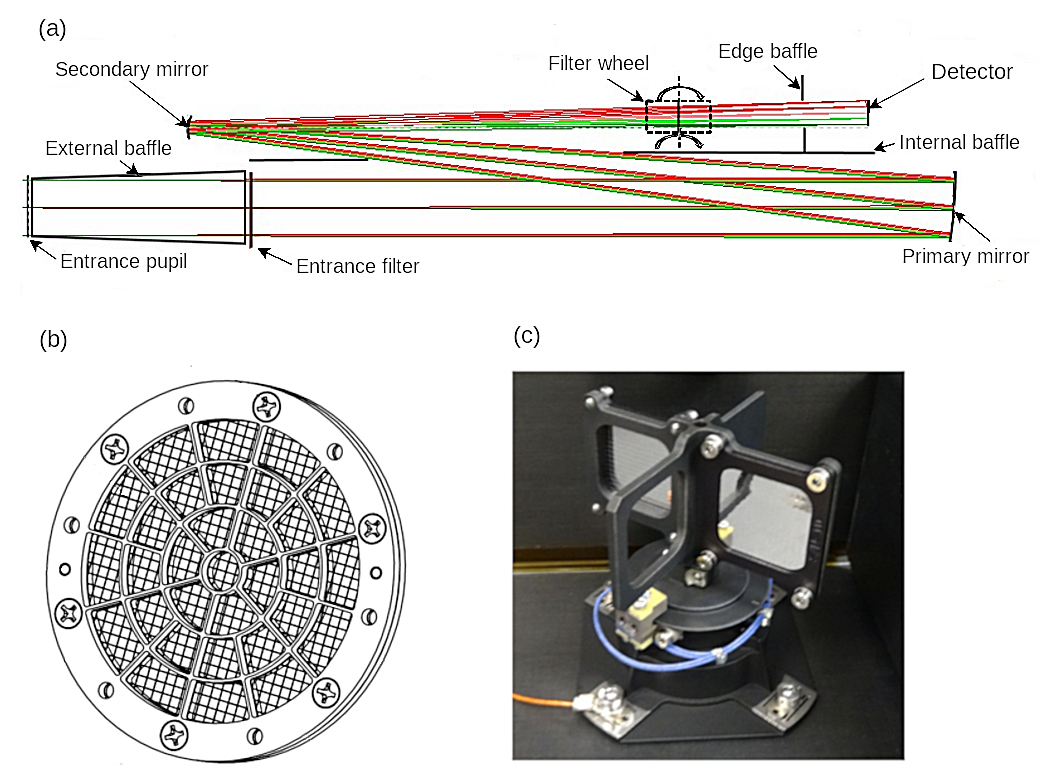}
    \caption{Schematics of \hrieuv. (a)~Instrument layout and optical light path, adapted from \citet{rochus2020}. (b)~Mounting of the entrance filter, provided by courtesy of the EUI team. (c)~Image of the filter wheel, reproduced from \citet{rochus2020} with permission from Astronomy \& Astrophysics, © ESO}
    \label{fig:schematics}
\end{figure}

\hrieuv\ is an off-axis Cassegrain telescope and is shown schematically in Fig.~\ref{fig:schematics}. It has an aperture of~\SI{47.4}{mm}. The optical entrance pupil of the system is roughly located at the position of the aperture. The primary mirror is located \SI{655.9}{mm} downstream from the aperture and has a diameter of \SI{66}{mm}, a radius of curvature of \SI{1518}{mm}, and a conic constant of~$-1$. The secondary mirror is located \SI{655.7}{mm} downstream from primary mirror and has a diameter of \SI{25}{mm}, a radius of curvature of \SI{257}{mm}, and a conic constant of~$-2.04$. The filter wheel is located \SI{306.7}{mm} downstream of the secondary mirror. Note that the filter wheel is not located at the exit pupil of the optical system, in contrast to the description in \citeauthor{rochus2020}(\citeyear{rochus2020}; private communication with the EUI team). Lastly, the detector is \SI{273}{mm} downstream from the filter wheel. The focal length of the entire optical system is~\SI{4187}{mm} \citep{rochus2020}. 

The incoming light arrives at the primary mirror, at an angle of \SI{3}{\degree} with respect to the optical axis of the mirror, and is reflected onward to the secondary mirror. This off-axis incidence angle introduces a slight anisotropy in the PSF. However, the effect could not be observationally resolved and was therefore neglected. Filters are located near the entrance pupil and in the filter wheel. These absorb visible light and off-band EUV radiation. Each filter is supported by a mesh grid, whose wires are aligned with the rows and columns of the detector pixels. The center-to-center wire spacings of the meshes, i.e., the mesh pitch, are \SI{1270}{\um}. The mesh wire widths are \SI{40}{\um}. For the entrance mesh, the number of illuminated wires in each direction is approximately $37$. For the mesh supporting the filters in the filter wheel, the number of illuminated mesh wires in each direction is approximately $16$.
The detector comprises a $3072 \times 3072$ pixel array with a pixel size of approximately $10 \times 10$~\si{\um^2}, corresponding to a physical size of $30.72 \times 30.72$~\si{\mm^2} and a plate scale of \SI{0.492}{\arcsec/px}. For scientific observations, only the central $2048 \times 2048$ pixels are used. Each detector pixel outputs a high-gain and a low-gain signal. The high-gain signal provides precise measurements with a low read noise but at the cost of a low saturation level. The low-gain signal provides a large dynamic range with high saturation levels but at the cost of higher read noise. The high-gain and low-gain signals can be recoded to a combined-gain signal, which then provides precise measurements over a large dynamic range. This recoding introduces a gain calibration factor between the low-gain and high-gain signal, resulting in an additional measurement uncertainty. All high-gain, low-gain, and combined-gain images are stored using lossless compression. For further details on the instrument, we refer to \citet{rochus2020}.

\section{The observations} \label{sec:observations}

\begin{figure}
    \centering
    \includegraphics[width=\textwidth]{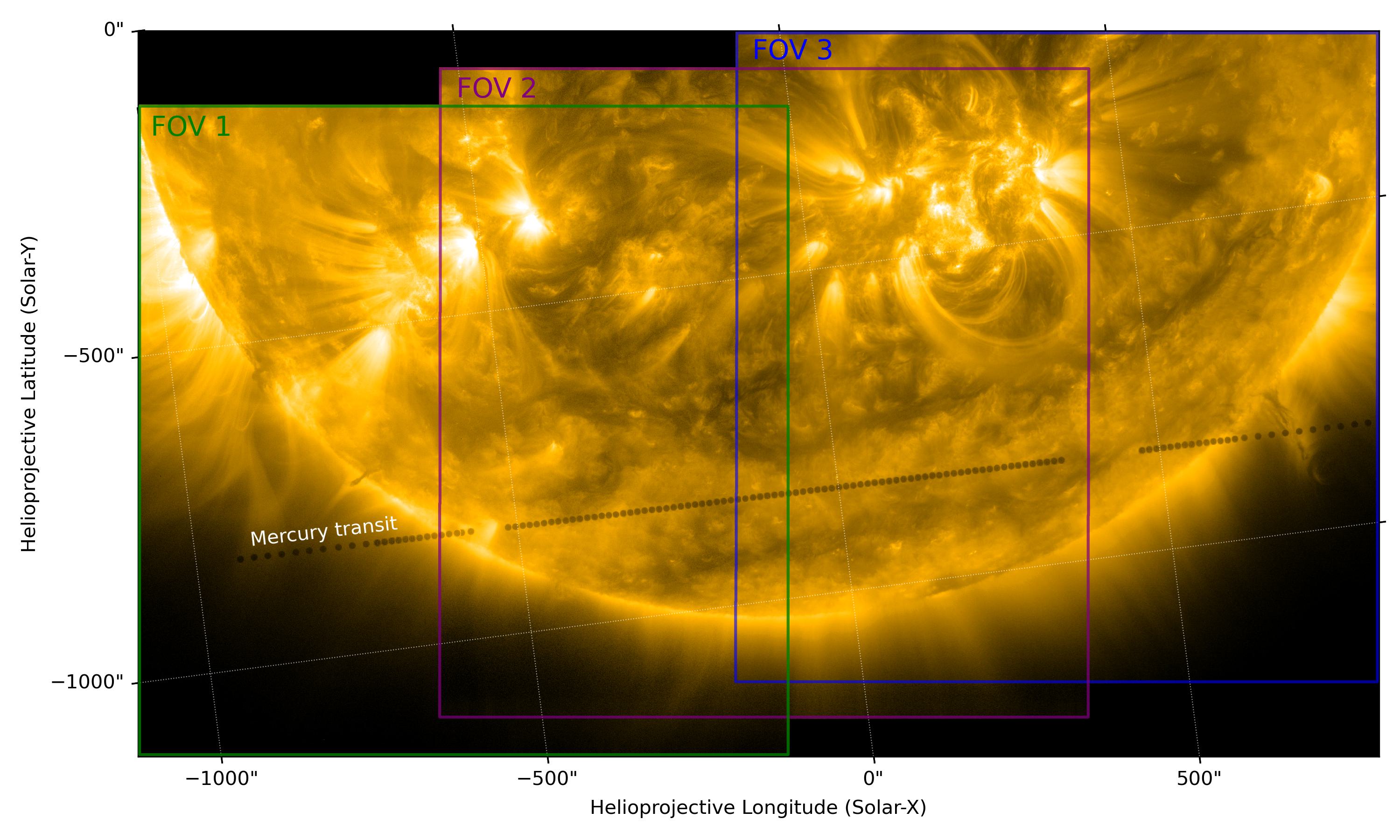}
    \caption{\hrieuv\ mosaic from Mercury transit observations on 2023~Jan~03. The dataset consists of~$126$ collected images during three pointings: the east limb (FOV~1), central meridian (FOV~2), and west limb (FOV~3). The mosaic also shows the transit locations of Mercury during all collected images. }
    \label{fig:Mercurytransit}
\end{figure}

\hrieuv\ observed the transit of Mercury on 2023~Jan~03, from 03:42~to~08:03 UT, capturing a total of $126$~relevant images. An overview of these observations is given in Figure~\ref{fig:Mercurytransit}. In each collected image, Mercury occulted a roughly circular area with a diameter of about $21.5$~pixels. The observations were conducted in three phases, each optimized to maximize the recorded signal from instrumental scattered light in the pixels occulted by Mercury.

The first phase followed Mercury’s ingress onto the solar disk. This field of view of \hrieuv, which we denote as FOV~1, was positioned over the eastern off-limb region of the Sun. From 03:42~to~04:11~UT, while Mercury was still far from the limb, \hrieuv\ acquired $10$~combined-gain images with $5$-second exposure times. As Mercury approached the solar limb between 04:13~and~04:35 UT, the instrument switched to $14$~high-gain images with shorter exposure times of $0.9$~seconds, to avoid large-scale saturation.
During the second phase, Mercury transited across the solar disk and the field of view was correspondingly shifted to the central meridian. We denote this as FOV~2. Between 04:44~and~06:52~UT, \hrieuv\ captured $78$~high-gain images with exposure times of $1.8$~seconds, covering the entire on-disk portion of the transit.
In the third and final phase, \hrieuv\ observed Mercury’s egress from the Sun into interplanetary space, with the field of view focused on the western off-limb region.  We denote this as FOV~3. From 07:11~to~07:32~UT, while Mercury remained close to the solar limb, $14$~high-gain images were taken using $0.9$-second exposures. As Mercury moved farther off-limb between 07:33~and~08:03~UT, \hrieuv\ resumed recording $10$~combined-gain images with $5$-second exposures.

We employ level~2 images of these observations. These have been re-processed by the EUI team with the most recent calibration routines as described at EUI's data release webpage\footnote{\url{https://doi.org/10.24414/z818-4163}}.  Additionally, for the combined-gain images, they performed a gain and offset calibration between the underlying low- and high-gain channels to bring them accurately onto the same intensity scale and further removed a large-scale gain gradient which was present across the images (private communication with the EUI team). We normalized the calibrated images by their exposure time. In the brightest regions, where long exposures caused saturation in the high-gain images, we replaced the affected pixels with values from the nearest-in-time, unsaturated combined-gain images. We analyze the resulting image intensities in digital numbers (DNs), maintaining the original detector orientations and pixel scale.

\section{A first guess on the amount of scattered and diffracted light} \label{sec:first_guess}

Small-scale occultations, such as the Mercury transit analyzed here, enable us to estimate a first guess for the amount of diffracted and scattered light directly from the observed intensities within the occulted regions. This approach provides a simple and fast assessment of the severity of instrumental scattering and diffraction without requiring a full PSF reconstruction.  In the following, we deduce this estimate from simple theoretical considerations and apply it to the observations. 

The observed signal in an image target pixel comprises the photons from the true image intensity distribution at that location and the photons that arise from the rest of the image and are instrumentally scattered and diffracted into the target pixel.  Thus, the observed signal can be written as
\begin{equation}
    I_\text{o,$\pmb{r}$}= \sum_{\pmb{\Delta} r\ge 0} I_\text{t,$\pmb{r}$-$\pmb{\Delta r}$} \ \text{psf}_{\pmb{\Delta r}} + \epsilon, \label{eq:psf}
\end{equation}
where   $I_\text{o,$\pmb{r}$}$ is the observed image intensity at the target location $\pmb{r}$;  $I_\text{t,$\pmb{r}$-$\pmb{\Delta r}$}$ is the true image intensity at a location $\pmb{r}$-$\pmb{\Delta r}$, i.e., the true intensity when no instrumental effects would be present; $\text{psf}_{\pmb{\Delta r}}$ is the amount of diffracted and scattered light from an image pixel into the target pixel in direction $\pmb{\Delta r}$, with  $\text{psf}_{\pmb{\Delta r}=0}$ corresponding to the amount of photons that are not scattered or diffracted; the sum is over the entire image plane, and $\epsilon$ is a noise component which we neglect in the following. 

When considering a homogeneously illuminated image of true intensity $\overline{I}_\text{t}$ and assuming no photon loss, i.e., $\sum_{\pmb{\Delta r}\ge 0} \text{psf}_{\pmb{\Delta r}} = 1$, the observed image intensity equals the true intensity, 
\begin{equation}
   I_\text{o,$\pmb{r}$} =  \overline{I}_\text{t}\ \sum_{\pmb{\Delta r}\ge 0} \text{psf}_{\pmb{\Delta r}} = \overline{I}_\text{t}.
\end{equation}
Since the true image is homogeneously illuminated, it follows that the intensities in the observed image are also likewise homogeneous, $\overline{I}_\text{o} = \overline{I}_\text{t}$. 

When we perturb this image by including a small occulting disk of radius $r_\text{disk}$, the true intensities at the location of the disk are zero. The sum then only goes over pixels outside of the disk, and the observed intensity in the disk center, $I_\text{o,dc}$, can be expressed by 
\begin{equation}
    I_\text{o,dc} = \overline{I}_\text{t}\ 
    \sum_{\pmb{\Delta r}>r_\text{disk}} 
    \text{psf}_{\pmb{\Delta r}}. 
\end{equation}
Introducing the small disk in the image has little effect on the observed intensities in the homogeneously illuminated portion of the image. It only results in no photons being scattered from the small area of the disk to the rest of the image, and therefore  $\overline{I}_\text{o} \approx \overline{I}_\text{t}$ remains approximately valid. It follows that 
\begin{equation}
   I_\text{o,dc} \approx \overline{I}_\text{o}\ 
   \sum_{\pmb{\Delta r}>r_\text{disk}} 
   \text{psf}_{\pmb{\Delta r}}.   
\end{equation}
The amount of light that is diffracted and scattered farther than the disk radius is then given by the ratio of $I_\text{o,dc}$ to ${\overline{I}}_\text{o}$ ,
\begin{equation}
    \sum_{\pmb{\Delta r}>r_\text{disk}} \text{psf}_{\pmb{\Delta r}} = \frac{I_\text{o,dc}} {{\overline{I}_\text{o}}}.
\end{equation}

\begin{figure}
    \centering
    \includegraphics[width=.7\textwidth]{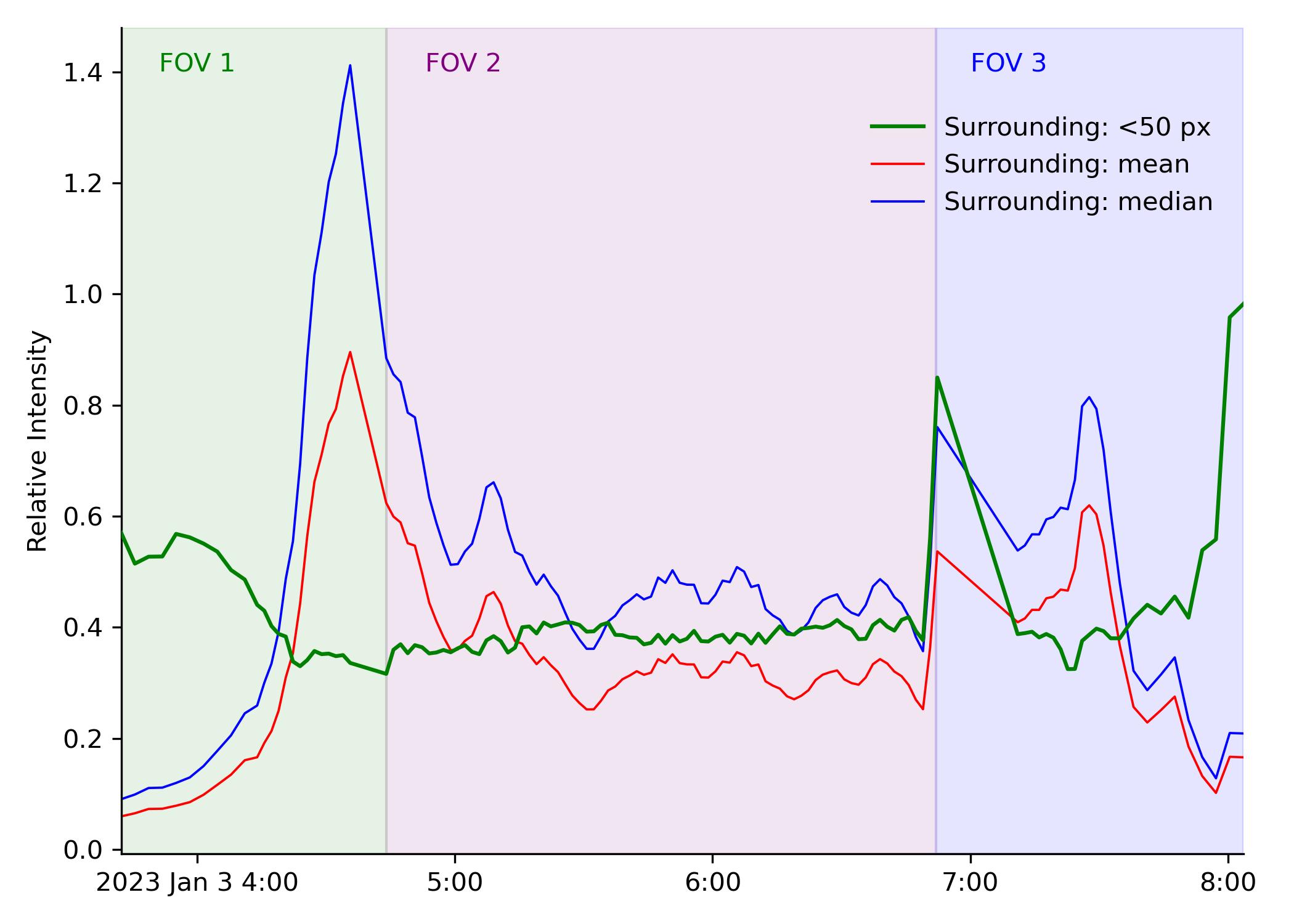}
   \caption{Intensity at Mercury’s center relative to its surrounding region. When the large-scale background can be approximated as roughly homogeneous, which is best satisfied near the center of FOV~2, this ratio directly measures the fraction of light that is scattered farther than the length of Mercury's radius, i.e., farther than \SI{10.75}{pixels}.} 
    \label{fig:first_guess}
\end{figure}

Figure~\ref{fig:first_guess} shows the transit evolution of this ratio between the observed intensity at the center of the Mercury occultation and the characteristic average intensity of the image. We quantified the characteristic average intensity ${\overline{I}_\text{o}}$ using three estimators: the mean intensity of the image, its median intensity, and the mean intensity within a 50-pixel-wide shell around Mercury. For the Mercury transit, the assumption of a homogeneously illuminated image is, at least statistically, best satisfied during the second phase of the observations, outlined by FOV~2 in Figure~\ref{fig:Mercurytransit}. In this phase, no exceptionally bright or dark features occurred near Mercury’s path apart from the beginning of the FOV~2 transit phase until 5:15~UT. In this phase, the image also exhibited no large-scale intensity gradient, except for the off-limb corona at the bottom of the image and an active region at the top right, which might create a small statistical bias, but are both reasonably far away from Mercury's transit path. Assuming that between 5:15 and 6:45~UT in FOV~2 the intensity in the intermediate surrounding of Mercury can be approximated as sufficiently homogeneous for our purposes, we find that the intensity ratio during this transit period is approximately \num{0.4 \pm 0.1}. Given that Mercury’s diameter corresponds to \SI{21.5}{pixels} and thus its radius is \SI{10.75}{pixels} in our observations, this implies that, as a rough estimate, about \SI{40\pm 10}{\percent} of light is diffracted and scattered to distances greater than \SI{10.75}{pixels}. This level of scattered and diffracted light is severe and thus requires a robust PSF modeling for correcting the recorded images.

\section{Diffraction from the entrance and filter-wheel mountings and meshes} \label{sec:diffraction}

Instrumental diffraction from the entrance aperture and filter-wheel mesh can be accurately modeled using the mechanical design specifications. We first summarize the theoretical concepts for diffraction modeling, then discuss their numerical implementation, and lastly analyze the resulting diffraction characteristics for \hrieuv.

\subsection{Theory}

We require three theoretical concepts: Fresnel far-field wave propagation, Fraunhofer near-field wave propagation, and diffraction from meshes. The following summary is based on the descriptions and formulas in \citet{bornwolf1999} and \citet{goodman2005}. 

\subsubsection{Fresnel propagation} \label{sec:fresnel}
The PSF of a telescope is given by the intensity of the complex wave field $E$ after propagation from the entrance pupil to the focal-plane detector,
\begin{equation}
\text{PSF} = \left| E_\text{detector} \right|^2.
\end{equation}
For \hrieuv, this requires modeling the propagation of the complex wave field from the entrance pupil, $E_\text{entrance}$, through the primary mirror, $E_\text{primary}$, the secondary mirror, $E_\text{secondary}$, the filter wheel, $E_\text{filterwheel}$, and finally to the detector in the focal plane, $E_\text{detector}$, 
\begin{equation}
 E_\text{entrance} \xrightarrow{\mathrm{Fresnel}} E_\text{primary} \xrightarrow{\mathrm{Fresnel}} E_\text{secondary} \xrightarrow{\mathrm{Fresnel}} E_\text{filterwheel} \xrightarrow{\mathrm{Fresnel}} E_\text{detector}.
\end{equation}

At telescope scales, wave propagation between optical components must be described using the Fresnel near-field formulation. Let $E_\text{1}$ denote the complex wave at a plane coinciding with one optical component, $E_2$ the wave at the subsequent optical component, $(x, y)$ the transverse coordinates in these planes, $z$ the propagation distance, and $\lambda$ the wavelength of the light. Then, the wave field at the plane at the second optical component is given by
\begin{align}
    E_2(x_2, y_2) &= \frac{\mathrm{e}^{\frac{i 2 \pi z}{\lambda} }}{i \lambda z} \iint E_1(x_1, y_1) \mathrm{e}^{\frac{i \pi}{\lambda z} \left((x_2 -x_1)^2 + (y_2 - y_1)^2\right)} dx_1 dy_1 \nonumber \\ 
   &= \frac{\mathrm{e}^{\frac{i 2 \pi z}{\lambda} } }{i \lambda z}\ \mathrm{e}^{\frac{i \pi }{\lambda z}(x_2^2 + y_2^2) } \ \mathcal{F} \left\{ E_1(x_1,y_1)\  \mathrm{e}^{\frac{i \pi }{\lambda z}(x_1^2 + y_1^2) } \right\} \left(\frac{x_2}{\lambda z}, \frac{y_2}{\lambda z} \right). \label{eq:fresnel}
\end{align}
In the second line of Equation~(\ref{eq:fresnel}), the Fresnel integral has been reexpressed by a Fourier transformation $\mathcal{F}$. The input field $E_1$ is first multiplied by a quadratic phase factor $\frac{i \pi }{\lambda z}(x_1^2 + y_1^2)$, then Fourier transformed and evaluated at the spatial frequencies $\left(\frac{x_2}{\lambda z}, \frac{y_2}{\lambda z} \right)$ , and finally multiplied by the quadratic phase factor $\frac{i \pi }{\lambda z}(x_2^2 + y_2^2)$ in the output plane.  

At each optical surface, the wave field is further modified.  At the entrance aperture and the filter-wheel mesh, portions of the wave field are blocked, which is modeled by applying amplitude masks with zero transmission in the obscured regions. At mirror planes, the wave field acquires an additional phase shift equal to twice the optical path difference between the mirror plane and the mirror sag,  
\begin{equation}
\text{phaseshift}\ (x, y) = \pm\ 2 \frac{2 \pi}{\lambda} \frac{\sqrt{x^2 + y^2}}{\mathrm{R}\ \left(1+ \sqrt{1- (1 + \mathrm{K}) \frac{x^2 + y^2)}{R^2}}\right) },
\end{equation}
where $\mathrm{R}$ is the mirror radius of curvature, $\mathrm{K}$ is the conic constant, and $(x, y)$ are coordinates relative to the mirror center. The plus sign applies to convex mirrors and the minus sign to concave mirrors. For off-axis designs, an additional phase term can be introduced to account for changes in propagation direction; however, this effect is neglected here for reasons discussed in the following sections.


The principal limitation of the Fresnel propagation is the computational cost. The computational array size of the discretized wave field at a given optical plane, $N_1 \times N_1$, is set by the physical extent of the optical component $X_1 \times Y_1$ and the spatial sampling $(dx_1, dy_1)$, 
\begin{equation}
    N_1 \times N_1 = \frac{X_1}{dx_1} \times \frac{Y_1}{dy_1}.\label{eq:sampling}
\end{equation}
The spatial sampling, in turn, determines the physical size of the propagated wave field at the subsequent downstream plane $X_2 \times Y_2$ through
\begin{equation}
X_2 \times Y_2 = \frac{\lambda z}{dx_1} \times \frac{\lambda z}{dy_1}. \label{eq:propsampling}
\end{equation}
Consequently, the spatial sampling $(dx_1, dy_1)$ and thus also the numerical sizes of the discretized wave fields are not a free parameters but are constrained by the optical geometry. Solving Equations~(\ref{eq:sampling}) and~(\ref{eq:propsampling}) for $N_1 \times N_1$ yields
\begin{equation}
    N_1 \times N_1 = \frac{1}{(\lambda z)^2}\ \left(X_1 X_2 \times Y_1 Y_2\right). \label{eq:arraysize}
\end{equation}
Therefore, the required computational array sizes scale linearly with the size of the optical components and inverse squared with the propagation distance. 

Applying this to the dimensions of \hrieuv, modeling Fresnel propagation from the entrance aperture to the primary mirror would require arrays of order $\num{300000}\times\num{300000}$, corresponding to a memory demand of approximately $12$~TB for a single Fourier transform. This exceeds currently available computational resources for a PSF derivation.  Here, we will perform a simplistic Fresnel wave propagation modeling in one dimension perpendicular to the optical axis instead of two, which enables us to test the validity of simplifications we describe in the following sections.  
 
 \subsubsection{Fraunhofer propagation} \label{sec:fraunhofer}
A simplification to the Fresnel propagation formulation is provided by the Fraunhofer formulation, which applies in the far-field regime of large propagation distances $z$. When
\begin{equation}
    z \gg R^2 / \lambda, 
\end{equation}
the quadratic phase term $(x_1^2 + y_1^2) / \lambda z$ in Equation~(\ref{eq:fresnel}) becomes negligible with~$(x_1, y_1) < R$, yielding the far-field expression, 
\begin{align}
    E_2(x_2, y_2) &= \frac{\mathrm{e}^{\frac{i 2 \pi z}{\lambda} } }{i \lambda z}\ \mathrm{e}^{\frac{i \pi }{\lambda z}(x_2^2 + y_2^2) } \ \mathcal{F}\left\{E_1(x_1, y_1)  \vphantom{\mathrm{e}^{\frac{i \pi }{\lambda z}(x_1^2 + y_1^2) }}\right\}\left(\frac{x_2}{\lambda z}, \frac{y_2}{\lambda z} \right). \label{eq:fraunhofer}
\end{align}
In most telescopes, the propagation distances are too small to reach the far-field regime. However,  there are two specific planes where the quadratic phase term introduced by Fresnel propagation is exactly canceled by the quadratic phase imparted by the telescope mirrors: wave propagation from the entrance pupil plane and from the exit pupil plane to the focal plane of the optical system \citep{goodman2005}. In these cases, we can neglect the intermediate optical components and model the wave propagation by the Fraunhofer formulation. The effective propagation distance for waves starting from the entrance pupil is the focal length of the system. For waves starting from the exit pupil, it is the distance between the exit pupil and the detector.

Applied to the \hrieuv\ entrance pupil, using the Fraunhofer approximation replaces the individual inter-component propagation distances of the Fresnel formulation with the full system focal length. As a result, the required computational array size, as given by Equation~(\ref{eq:arraysize}), is reduced to approximately $\num{14500}\times\num{14500}$, corresponding to a memory demand of about 26~GB and making the computation tractable.

\subsubsection{Analytical solution for mesh diffraction} \label{sec:meshanalytical}
For diffraction by a regular mesh, an analytical solution for the far-field intensity distribution is given by
\begin{equation}
I (x, y) = \sinc^2\left(\frac{\pi w x}{ \lambda z}\right)\  \left(\frac{\sin\left(N_\text{mesh} \pi d x/ (\lambda z)\right)}{\sin\left(\pi d x/ (\lambda z)\right)} \right)^2 \ \sinc^2\left(\frac{\pi w y}{\lambda z}\right)\  \left(\frac{\sin\left(N_\text{mesh} \pi d y/ (\lambda z)\right)}{\sin\left(\pi d y/ (\lambda z)\right)} \right)^2, \label{eq:mesh_diffraction}
\end{equation}
where  $d$ is the mesh pitch, $w$ is the mesh window width (defined by the mesh pitch minus the mesh wire width), and $N_\text{mesh}$ is the number of mesh windows. We use the definition of $\sinc(a) = \sin(a)/a$, where $a$ is an unspecified variable. The spacing of the diffraction peaks is set by $\lambda / d$, while the envelope of the diffraction pattern is controlled by $\lambda / w$. A larger mesh pitch $d$ produces diffraction peaks that are closer together, resulting in a more compact diffraction pattern. A larger window width $w$ concentrates the diffracted energy more narrowly near the center, yielding a more compact overall diffraction pattern. 

To accurately sample the intensity distribution of the diffraction pattern, the spatial sampling must resolve the individual diffraction peaks. The full width at half maximum $\delta$  of the diffraction peaks is approximately given by 
\begin{equation}
    \delta \approx \frac{0.885\ \lambda z}{N_\text{mesh}\,d}. \label{eq:widthpeaks}
\end{equation}
For diffraction by the \hrieuv\ filter-wheel mesh, evaluated at the detector, this condition requires a detector-plane sampling of at least \SI{0.08}{\um}, which is too small to be efficiently computationally evaluated in two dimensions. However, since Equation~(\ref{eq:mesh_diffraction}) is separable in $x$ and $y$, the diffraction pattern can be computed by first evaluating the one-dimensional diffraction patterns in each direction at high resolution, rebinning them to the detector sampling, and then forming the two-dimensional pattern as the outer product of the two one-dimensional results. Details of this numerical implementation are given in \citet{hofmeister2025}.

\subsection{Implementation} \label{sec:diff_implementation}

\begin{figure}
    \centering
    \includegraphics[width=\textwidth]{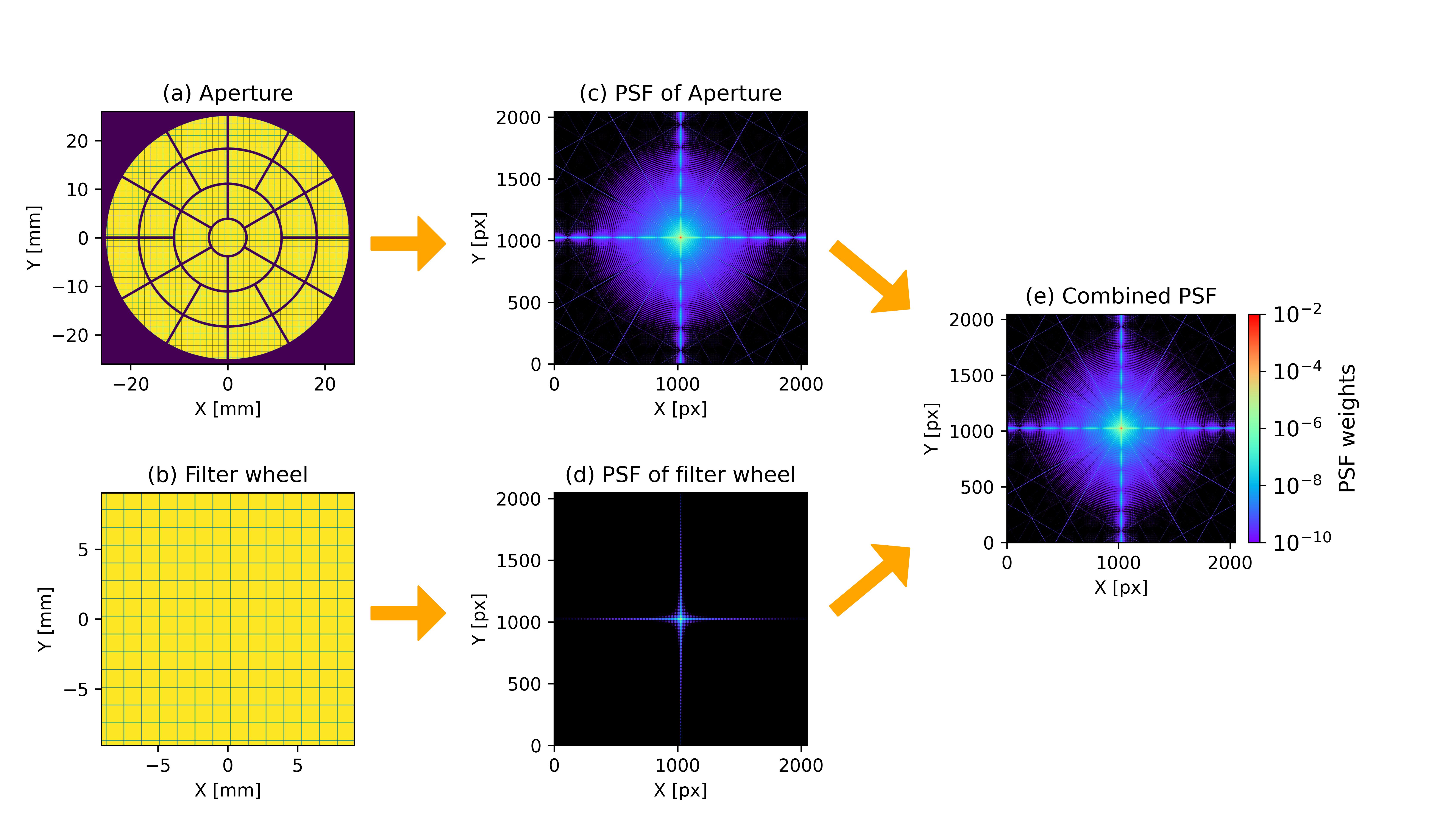}
    \caption{Derivation of the PSF describing the diffraction pattern, using the simplified-combined method described in Section~\ref{sec:diff_implementation}. (a)~and~(b) Amplitude masks describe the mountings and meshes of the aperture and filter wheel, respectively. Yellow indicates open areas and dark purple closed ones. (c) For the aperture, the incoming wave field is propagated to the detector using Fraunhofer propagation to produce the diffraction pattern. (d) For the filter wheel, the diffraction pattern is computed using the analytical solution for diffraction on a mesh. (e) The combined PSF is computed by the convolution of the two individual diffraction patterns. } 
    \label{fig:diffraction_derivation}
\end{figure}

\begin{figure}
    \centering
    \includegraphics[width=\textwidth]{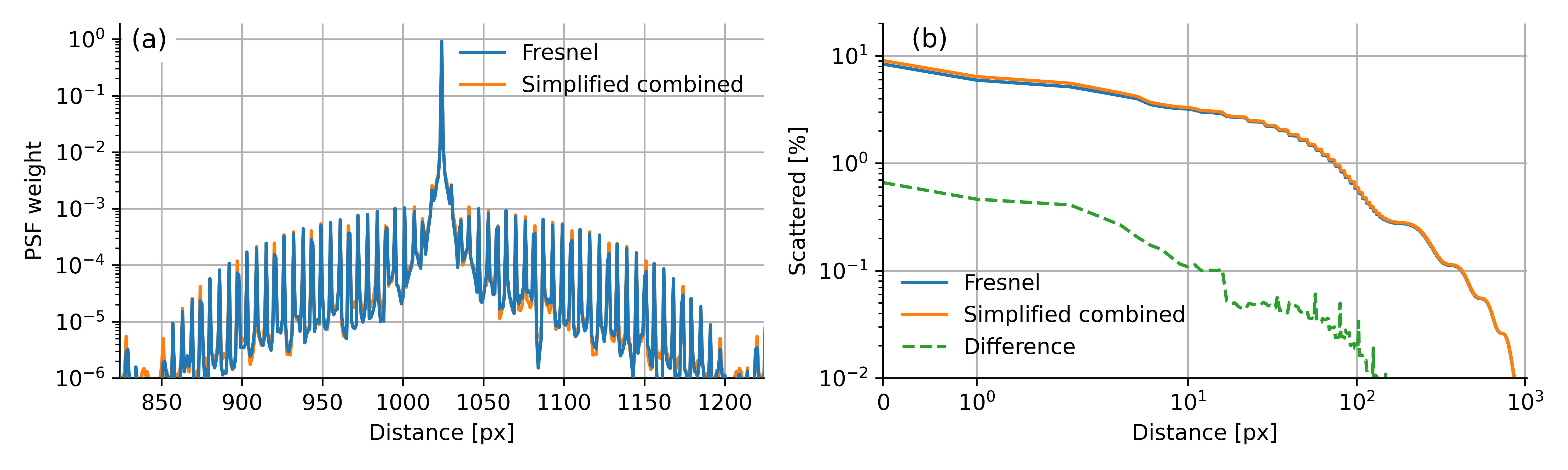}
    \caption{PSF of \hrieuv\ as derived in one dimension. (a) PSF weights vs.\ detector location, zoomed in to the PSF center. (b) Percentage of light that is scattered farther than a given distance vs.\ distance. Blue lines are computed by Fresnel wave propagation and orange lines by the simplified-combined method. The green line in panel~(b) gives the difference between the two modeling methods.  } 
    \label{fig:diffraction_1d}
\end{figure}

Physically, the PSF of \hrieuv\ should be obtained by propagating the complex wave field from the entrance pupil to the detector using Fresnel propagation. This is required because the filter-wheel mesh is located within the optical path rather than at the exit pupil, which precludes straightforward simplifications. However, full two-dimensional Fresnel propagation is computationally not feasible. We therefore seek an approximate solution that is computationally tractable while remaining physically accurate.

A natural approach is to try to separate the diffraction effects of the entrance filter and the filter-wheel mesh. Diffraction from the entrance filter can be modeled using Fraunhofer propagation from the entrance pupil to the detector (Section~\ref{sec:fraunhofer}), while diffraction from the filter-wheel mesh to the detector can be described by the analytical solution presented in Section~\ref{sec:meshanalytical}. The resulting PSF can then be approximated by convolving the individual diffraction patterns in the detector plane. This approach is illustrated in Figure~\ref{fig:diffraction_derivation} and is referred to in the following as the simplified-combined solution.

The main caveat of this simplification is whether it is physically justified. The analytical solution for mesh diffraction assumes the far-field regime, whereas the detector is is clearly located in the near field of the filter-wheel mesh. In addition, combining the diffraction patterns from the entrance filter and the filter wheel by convolution is only valid if their respective diffraction peaks do not significantly interfere, that is, if they are physically well separated. As we are not aware of analytical tests for these assumptions, we investigate their validity by modeling the telescope in one transverse dimension.

In Figure~\ref{fig:diffraction_1d}(a), we show the resulting one-dimensional PSF obtained from Fresnel modeling of the telescope following Section~\ref{sec:fresnel}, compared to the simplified-combined solution. Both methods reproduce the same overall behavior, although the diffraction peaks in the Fresnel solution are slightly lower than those of the simplified-combined solution. This difference arises because diffraction from the filter-wheel mesh is still in the near-field regime. At very high numerical resolution, the Fresnel solution reveals that the apparent diffraction peaks are in fact slightly extended wave oscillations, i.e., the wave interference caused by the mesh and observed at the detector has not yet fully converged into distinct far-field diffraction peaks. This difference in the modeling methods, however, becomes small at the comparably coarse resolution of the detector pixels, resulting in this good overall match. 

In Figure~\ref{fig:diffraction_1d}(b), we compare both methods with respect to the total energy diffracted to large distances in the detector plane. The two approaches again show a good agreement, with absolute differences in the diffracted energy below \SI{1}{\percent}. Based on this one-dimensional analysis, we conclude that the simplified-combined solution provides an adequate approximation and therefore apply it to the full two-dimensional case.

\subsection{Results}
\begin{figure}
    \centering
    \includegraphics[width=\textwidth]{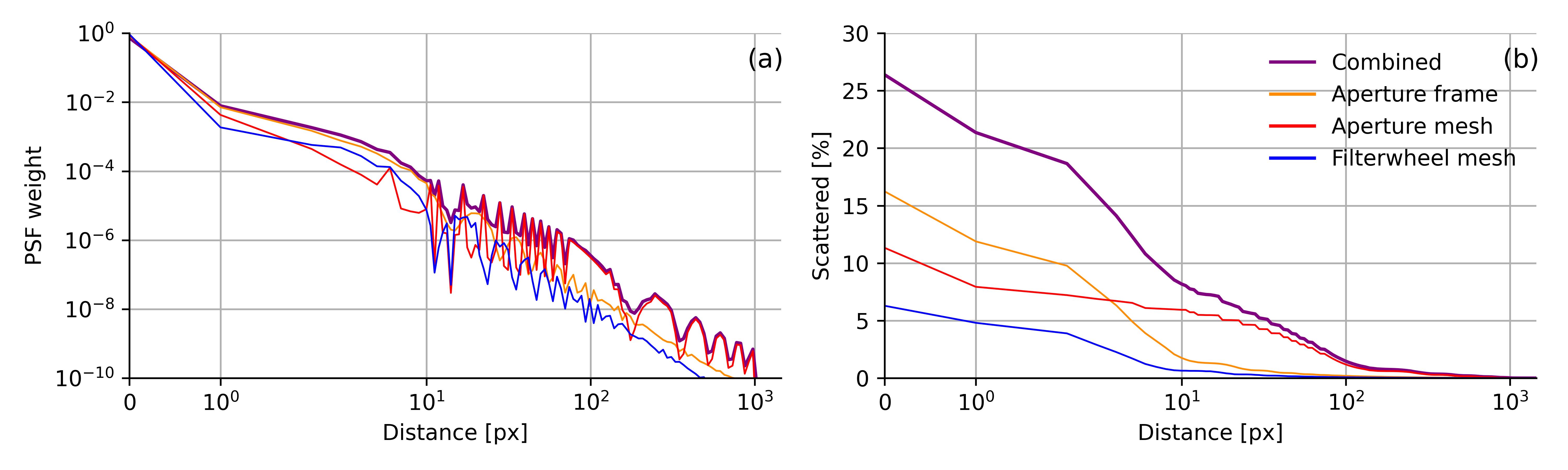}
    \caption{PSF describing the diffracted light of the combined system and of the individual components. (a) PSF weights vs.\ distance from PSF center. (b) Percentage of light that is scattered farther than a given distance.} 
    \label{fig:diffraction_results}
\end{figure}

We computed the two-dimensional \hrieuv\ diffraction PSF using the simplified-combined approach, which is shown in Figure~\ref{fig:diffraction_derivation}(e). The PSF is dominated by diffraction peaks from the mesh supporting the entrance filter, with diffraction peak spacings of $5.74$~pixels in the horizontal and vertical directions. When zooming into the PSF, concentric rings and a web-like structure become visible.  These arise from diffraction on the entrance-filter mounting. Conversely, the diffraction peaks from the filter-wheel mesh cannot be spatially resolved. 

In Figure~\ref{fig:diffraction_results}(a), we show this PSF, as well as the PSFs of the individual diffracting components, namely the entrance-filter mesh, the entrance-filter mounting, and the filter-wheel mesh, as a function of radial distance from the PSF center. The corresponding cumulative fraction of light diffracted beyond a given distance is shown in panel~(b). Taken together, the meshes and mountings diffract approximately \SI{26}{\percent} of the incident light to large distances, extending over several hundred pixels. The dominant contribution to the medium- and long-range diffraction arises from the mesh of the entrance filter. At smaller radii, up to about 10~pixels, the entrance-filter mesh and mounting as well as the filter-wheel mesh all produce noticeable diffraction features. However, the components diffract only about \SI{8}{\percent} of the total light beyond \SI{11.5}{pixels}, which is far too small to explain the \SI{\sim 40}{\percent} of scattered and diffracted light observed near the center of the Mercury occultations (Section~\ref{sec:first_guess}). We therefore conclude that an additional source of scattering must be present in the instrument.

We validated the computed diffraction PSF by comparing the orientation of the main diffraction arms and the spacing of the corresponding diffraction peaks with those observed during seven strong flare events. The computed diffraction PSF reproduce the observed orientations of the diffraction pattern to within \SI{0.3}{\degree}. The spacings of the computed PSF diffraction peaks to those observed agree within \SI{0.02}{pixels}, which corresponds to an error in the computed spacings of \SI{<0.3}{\percent}. A further fine-tuning of the assumed mesh window width and pitch was not feasible, as the flare observations revealed diffraction peaks only within the first diffraction order. A more precise determination would have required substantially stronger flares, producing multiple higher diffraction orders detectable across larger regions of the detector.

\section{The diffuse scattered light} \label{sec:diffuse}

In EUV imaging instruments, it is generally assumed that a significant fraction of the incoming light is scattered by the micro-roughness of the mirrors, producing diffuse scattering over large distances. This scattering is likely the primary cause for the scattered light we observe in the Mercury occultations. In the following, we use the observed intensities in the Mercury occultations to reconstruct the spatial profile of the scattered light, following closely the methodology of \citet{hofmeister2025}. In this section, we first describe the inversion method, then present the results, and lastly perform a short evaluation of the robustness and accuracy of the results.

\subsection{The inversion algorithm}

\begin{figure}
    \centering
    \includegraphics[width=\textwidth]{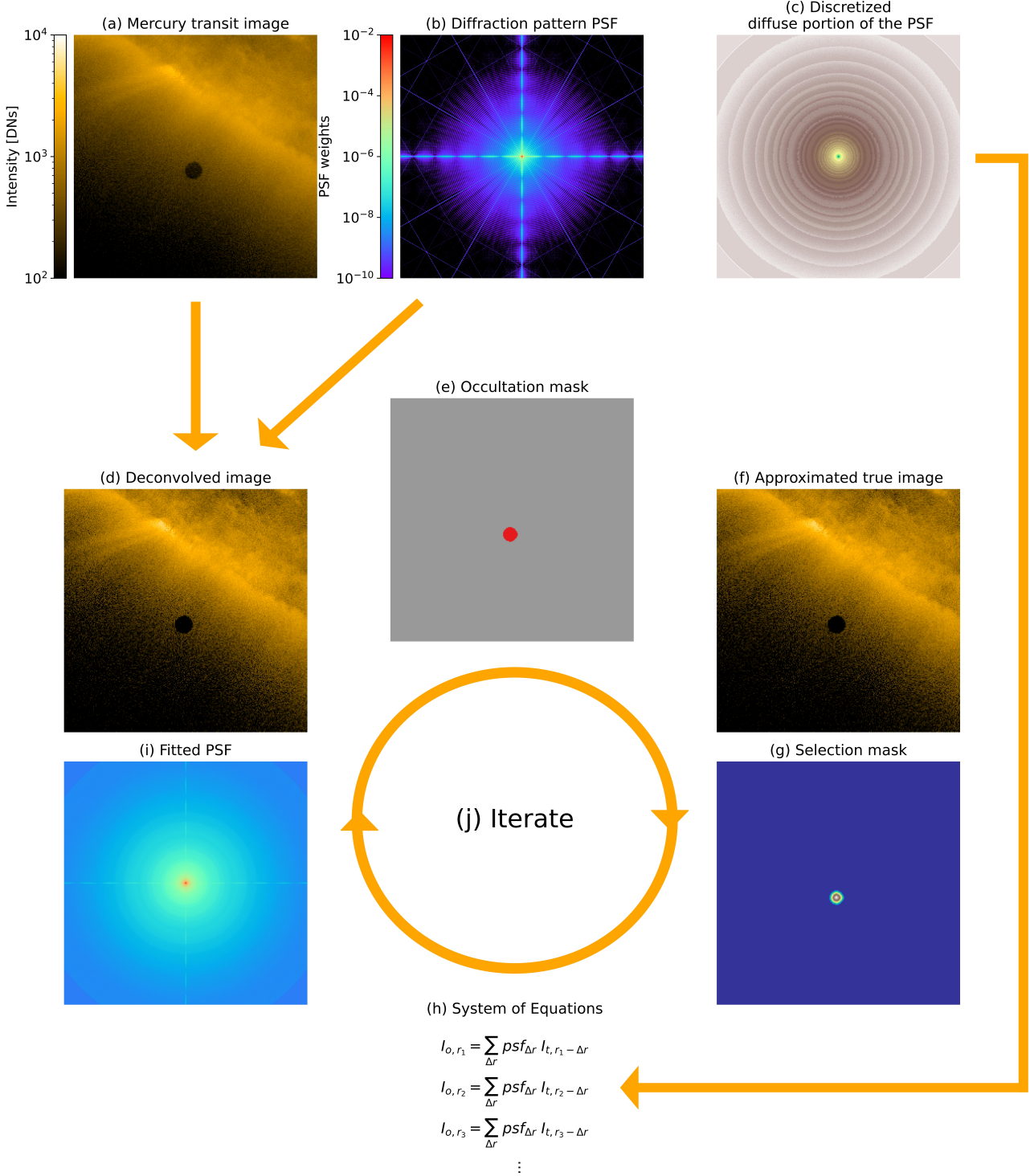}
    \caption{ Flowchart illustrating the calibration process for the diffuse scattering component of the PSF. The inputs consist of (a)~Mercury transit images, (b)~the known portion of the PSF representing the diffraction pattern, and (c)~a spatial segmentation of the PSF, where each segment corresponds to one PSF coefficient to fit. From these, we first compute (d) a PSF-deconvolved image and (e) a mask marking the pixels occulted by Mercury. Setting the values of the occulted pixels in the deconvolved image to zero yields (f) an initial estimate for the true image. We then select specific occulted pixels, which we use to fit the scattered light component, and call this mask (g) the Selection Mask. Using the PSF segmentation from (c), the approximate true image from (f), and the Selection Mask from (g), we construct a system of equations, which describes (h) the scattered-light contribution. Solving this system of equations provides an estimate for the scattering PSF. Combining the scattering PSF with the diffraction PSF results in (i) the total PSF. The total PSF is then used to refine the estimate of the true image, which in turn provides a refined PSF. This iterative process~(j) continues until the solution for the PSF and for the approximated true image converges.}
    \label{fig:tail_overview}
\end{figure}

The intensity in the occulted pixels during the Mercury transit contain information on the distribution of the scattered light: when Mercury is far off limb within the dim interplanetary space background, the scattered light in the occulted pixels is dominated by long-distance scattered light from the solar disk. Conversely, when Mercury is close to a bright region on the solar disk, such as an active region or the solar limb, the scattered light is dominated by short-distance scattered light from these nearby bright regions. Therefore, by sampling the scattered light in the occulted pixels during Mercury's passage over the solar disk, one can reconstruct how light is redistributed as a function of scattering distance, which is equivalent to the instrumental PSF.

For the PSF reconstruction from the Mercury transit images, we employ the inversion algorithm of \citet{hofmeister2022}. The inversion is based on a modified form of the PSF-defining equation,
\begin{align}
I_\text{o,$\pmb{r}$} &= \sum_{S} \widetilde{\text{psf}}_S \sum_{\pmb{\Delta r} \in S} I_\text{t,$\pmb{r}$-$\pmb{\Delta r}$}  
+ \left( 1 - \sum_S n_S\,\widetilde{\text{psf}}_S\right) 
\sum_{\pmb{\Delta r}} I_\text{t,$\pmb{r}$-$\pmb{\Delta r}$}\,\overline{\text{psf}}_{\pmb{\Delta r}} + \epsilon \\ 
&= \sum_{S} \widetilde{\text{psf}}_S\,A_S  
+ \left( 1 - \sum_S n_S\,\widetilde{\text{psf}}_S\right) B, 
\label{eq:psf_px3}
\end{align}
where $I_\text{o,$\pmb{r}$}$ is the observed intensity at a given pixel position $\pmb{r}$ and $I_\text{t}$ is the true intensity that would be measured without instrumental effects. The parameters $\widetilde{\text{psf}}_S$ are the PSF coefficients for which we will fit; $\overline{\text{psf}}$ represents known PSF contributions, such as from the diffraction pattern; $S$ denotes segments in the spatially segmented PSF each containing $n_S$ pixels; $\pmb{\Delta r}$ is the distance vector from the PSF center in pixels; and $\epsilon$ is a noise term. In the second line of Equation~(\ref{eq:psf_px3}), we have merged all known quantities that can be derived from the image into the coefficients $A_S$ and $B$, as detailed below.

Equation~(\ref{eq:psf_px3}) holds for each image pixel, forming a large system of equations. Within this system of equations, the observed intensities are all known and the true intensities are known to be zero within the occulted pixels. The true intensities outside the occultation can be estimated simultaneously with the PSF by an iterative approach: one fits alternately the PSF using an approximate true image and then updates this true image using the fitted PSF. The unknown PSF coefficients $\widetilde{\text{psf}}_S$ can thereby be determined by a multilinear fit on the system of equations, provided that the selected PSF segmentation is appropriate, the number of fit coefficients is smaller than the number of unique lines in the system of equations, and the noise $\epsilon$ is statistically negligible. The a-priori information that the true intensity in the occulted pixels is zero makes this semi-blind approach robust  \citep{hofmeister2022}. 

In practice, the method proceeds via the following steps, which are illustrated in Figure~\ref{fig:tail_overview}:  
(a)~collect appropriate Mercury transit images;  
(b)~compute the known diffraction PSF;  
(c)~spatially discretize the to-be-fitted PSF component into segments;  
(d)~deconvolve the images;  
(e)~identify the occulted pixels;  
(f)~approximate the true image;  
(g)~sample the occulted pixels;  
(h)~set up the system of equations;  
(i)~fit the PSF coefficients; and  
(j)~iterate steps (d)--(i), until both the PSF and the approximated true image have converged. 
In the following, we describe each of these steps in more detail.

{\it (a) Collect appropriate Mercury transit images -- }
We use the $126$ images recorded during the 2023~Jan~03 Mercury transit, which have been described in Section~\ref{sec:observations}.  As Mercury transits are rare, this observing campaign was specifically designed to obtain high-quality data for our PSF reconstructions. The observations began with Mercury’s ingress far off the eastern solar limb, continued throughout its transit across the solar disk, and extended through its egress far off the western solar limb, thereby sampling the scattered-light profile as a function of distance from the solar disk and bright solar features. In these observations, the exposure times were maximized to increase the signal in the occulted pixels while keeping Mercury’s motion below one pixel. Due to the long exposures, we accepted that some bright regions might saturate. These saturated regions have been re-filled with their observed intensities from the closest-in-time non-saturated images. An example of a zoomed-in Mercury observation is shown in Figure~\ref{fig:tail_overview}(a).

{\it (b) Compute the diffraction pattern PSF -- }
The inversion algorithm enables one to incorporate a known component of the PSF, for instance from theoretical calculations, and fit only the remaining part required to reproduce the observed scattered light. As diffraction in \hrieuv\ is well characterized by the mechanical drawings from the mountings and meshes holding the filters, we used the diffraction portion of the PSF derived in Section~\ref{sec:diffraction} as the fixed PSF component in Equation~(\ref{eq:psf_px3}), shown in shown in Figure~\ref{fig:tail_overview}(b).

{\it (c) Discretize the diffuse PSF component -- }
We discretized the PSF spatially into segments. Each PSF segment corresponds to one PSF coefficient to fit. Because the off-axis angles of \hrieuv\ are small, we assumed the scattering to be nearly isotropic and therefore used concentric shells as the structuring elements for PSF segmentation. Since the scattered light intensity typically decreases steeply, often following an approximate power law, the PSF core, where the intensity gradient is strongest, must be sampled at higher spatial resolution than the PSF tail, where variations are more gradual. Therefore, we discretized the PSF into 50 concentric shells: 10 shells within the first $5$~pixel radius, $5$ additional shells up to a radius of $10$~pixels, and $35$~shells with exponentially increasing widths out to the edge of the PSF. This discretized PSF is shown in Figure~\ref{fig:tail_overview}(c). To test for possible anisotropy, we also performed the inversion assuming direction-dependent scattering, where we divided each shell into 12 angular segments with angular widths of \SI{15}{degrees}. However, as shown in Section~\ref{subsec:diffuse_eval}, we did not find any significant anisotropies. 

{\it (d) Deconvolve the Mercury images -- }
To increase the accuracy of the recorded images, we deconvolved the images with the current best approximation of the PSF. In the first iteration, this is the diffraction PSF. In subsequent iterations, the diffraction PSF is combined with the fitted PSF from the previous iteration, containing the most recent guess on the diffuse scattered light. For the deconvolution, we used the Richardson-Lucy algorithm with a fixed number of $20$~iterations, which provides high-quality results even for the far off-limb, low signal-to-noise regions \citep{richardson1972, lucy1974}. An example of a deconvolved image is shown in Figure~\ref{fig:tail_overview}(d).

{\it (e) Identify occulted pixels -- }
To identify the pixels occulted by Mercury in a given transit image, we first generated a proxy image of the underlying solar emission by averaging the two images taken closest in time. We then simulated the occultation by applying a circular mask representing Mercury’s size and approximate position, setting the corresponding pixels in the proxy image to zero, and convolving the result with the current PSF approximation to account for scattered light. By adjusting the mask position and comparing the simulated image with the observation, we accurately determined Mercury’s location. Finally, the resulting occultation region was contracted by one pixel to exclude partially occulted pixels. An example of such a derived occultation mask is shown in Figure~\ref{fig:tail_overview}(e).

{\it (f) Approximate the true image -- }
To approximate the true image, we use the PSF-deconvolved image, which provides a good representation for the true image for the illuminated regions, and set the intensities within the occulted area to zero. This is shown in Figure~\ref{fig:tail_overview}(f).

{\it (g) Sample the occulted pixels and (h) set up the system of equations -- }
To constrain the PSF, we combined information from all $126$ Mercury transit images. The long-distance scattered light is sampled by the evolution of the intensities in the occultations during the Mercury transit, which cover a range of distances from the solar disk and bright features in the field of view. The short-distance scattered light is sampled from the individual occultations as a function of distance from the occultation edge. To better balance the dataset with respect to distance from the occultation edge, we assign the occulted pixels to one-pixel-wide concentric shells within the Mercury occultation masks. We utilize these shells to later resample the dataset by drawing an equal number of pixels from each shell, which improves the representation of the radial intensity decrease toward the planet’s center. We refer to this set of concentric shells as the Selection Mask, shown in Figure~\ref{fig:tail_overview}(g). To further improve the fit for very short scattering distances across Mercury's occultation edge, we also included ten randomly selected unocculted pixels within $5$~pixels of the occultation edge from each image. Altogether, this yielded \num{51660} pixels, corresponding to \num{51660} lines in the system of equations. For each line, we calculated the coefficients $A_S$ and $B$ of Equation~(\ref{eq:psf_px3}). Finally, we brought the observed intensities to the same scale by normalizing each line by its observed intensity. This step is required because the intensities in the occulted pixels vary by more than one order of magnitude depending on Mercury’s position during transit.

{\it (i) Fit the PSF from the system of equations -- }
With $I_\text{o,$\pmb{r}$}$, $A_S$, and $B$ known in the system of equations defined by Equation~(\ref{eq:psf_px3}), we determined the diffuse PSF coefficients $\widetilde{\text{psf}}_S$ via a multilinear fit.
During the first three iterations of the algorithm, the PSF coefficients were fitted directly, without any parameterization. We selected \num{1100} equations from the system, i.e., \num{1000} stratified samples from the occulted pixels, ensuring that each concentric shell in the Selection Mask contributes an equal number of pixels, and \num{100} samples from the unocculted pixels near Mercury’s boundary. Choosing more equations degraded the fit, as the limited number of central pixels from all Mercury occultations made the solution prone to overfitting. The PSF coefficients were then obtained from this system of equations using the Trust Region Reflective (TRF) algorithm, a well-established method for large-scale bound-constrained optimization \citep{branch1999}. This sampling and fitting was repeated \num{100} times; the \num{30}~least accurate fits were discarded, and the remaining \num{70}~fits were averaged to obtain a reliable estimate.

In the following three iterations, the fit coefficients were parameterized using a softened power law,
\begin{equation}
\widetilde{\text{psf}}_S(\Delta r) =
a\,\frac{1}{\left(b + \Delta r\right)^c},
\label{eq:psf_fit_tail_wing}
\end{equation}
as justified in the following section. This parameterization strongly reduced the number of fit parameters to counteract overfitting, leading to a robust result.
The fitted diffuse PSF were finally combined with the diffraction PSF to obtain the total PSF, shown in Figure~\ref{fig:tail_overview}(i).

{\it (j)~Iterate -- }
The procedure was iterated six times from~(d) to~(i). In the first three iterations, the unparameterized fit converged, providing an initial estimate of the total PSF. 
In the following three iterations, the refined fit using the soft power law function converged, providing the final PSF solution.

\subsection{Results}

\begin{figure}
    \centering
    \includegraphics[width=\textwidth]{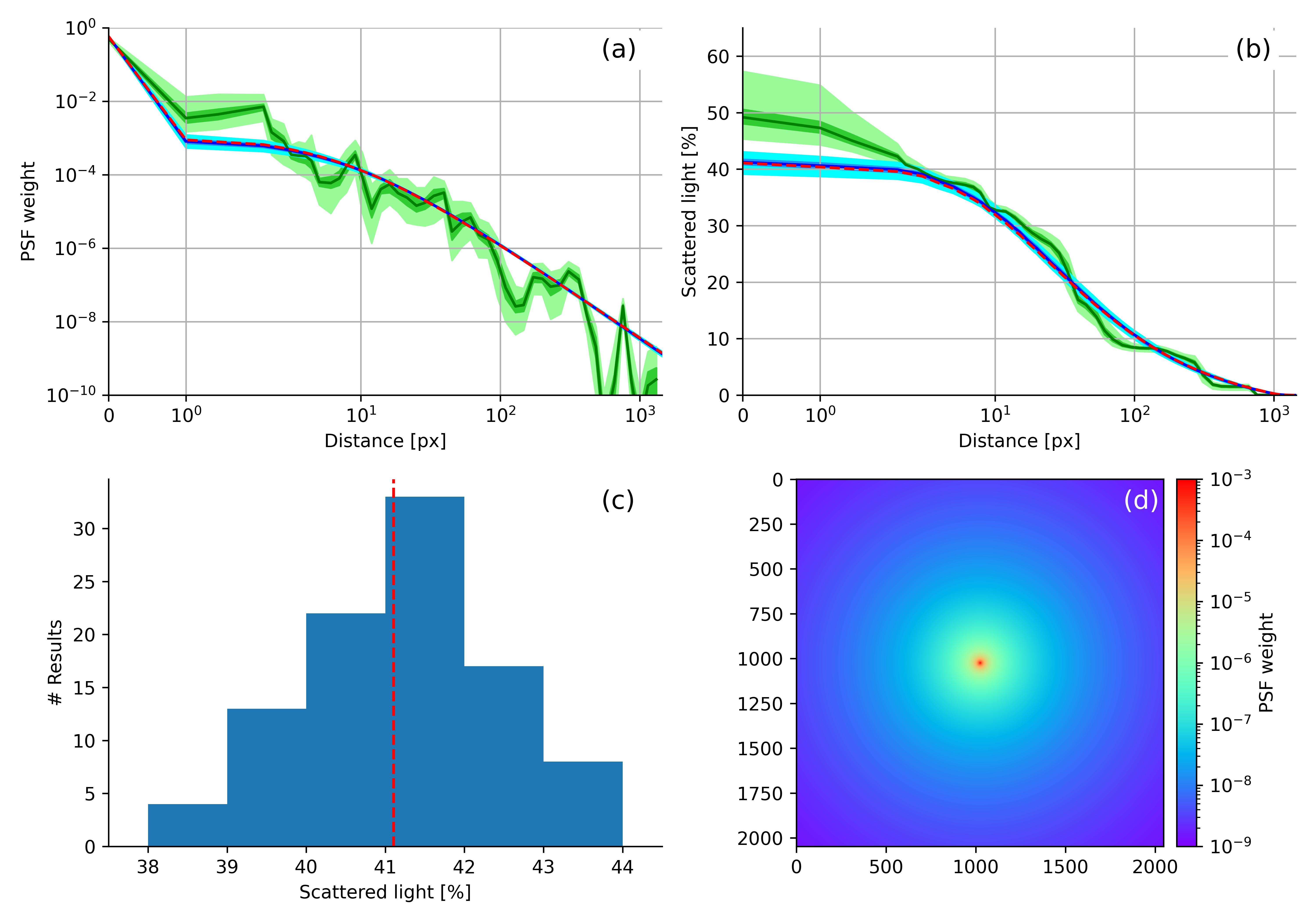}
    \caption{Results for the diffuse scattered light. (a)~The fitted PSF weights as function of distance from the PSF center, and (b)~the corresponding amount of light that is scattered farther than a given distance. The green and blue lines and bands show the median, $1\,\sigma$, and $2\,\sigma$ confidence levels for the non-parametric (green) and softened power-law (blue) fits, respectively. The red dashed line shows the final fit result as derived from the entire dataset. (c)~Histogram of the total amount of scattered light, as derived from the uncertainty analysis for the soft-power-law fit. The amount of scattered light for the final fit result is marked in red. (d)~Final PSF for the diffuse scattered light.}
    \label{fig:tail}
\end{figure}

We fitted the PSF for \hrieuv\ using all available Mercury transit images. This section presents the diffuse scattered-light component of the PSF as obtained from the inversion algorithm; the complete PSF is shown in Section~\ref{sec:finalPSF}. To assess the inversion uncertainty, we performed ten iterations of a tenfold cross-validation on temporally-stratified random samples of the dataset, as we now describe. First, the data were divided into ten temporally sequential blocks. 
Nine blocks were used for fitting and the remaining block for evaluation, with the evaluation block rotated through all ten blocks so that each served once as the evaluation set. To reduce cross-talk between fitting and evaluation, we removed in each run the three images closest to the evaluation block from each adjacent fitting block. Then, we randomized the images in each fitting block by randomly re-selecting images from the block, allowing repeated selections. We fitted the PSF on the images from the fitting blocks and subsequently evaluated the result on the images from the evaluation block. Repeating this procedure for each of the ten evaluation blocks across ten randomized runs yielded $100$ PSF solutions. From these solutions, we derived the median solution and the corresponding $1\,\sigma$ and $2\,\sigma$ uncertainties.

In Figure~\ref{fig:tail}, we present the inversion results. Panel~(a) shows the fitted PSF weights for the diffuse scattered-light component and panel~(b) the corresponding amount of light that is scattered farther than a given distance. The green lines represent the results from the unparameterized fit, the blue lines the soft–power-law fit derived from the uncertainty analysis, and the red lines the final fit obtained from the full dataset.

The fitted unparameterized PSF weights follow an overall power-law dependence on scattering distance, but exhibit pronounced oscillations. This behavior likely indicates overfitting: the Mercury transit dataset contains too few independent samples of occulted pixels to suppress small calibration errors across the images. As a result, the PSF fit can partially adjust to these errors, whose average is likely non-zero for each PSF scattering distance, leading to the observed fluctuations. However, since these calibration errors are expected to be small, the individual fit solutions can be assumed to oscillate around the true PSF. To regularize the fit, we tested several parameterizations of the PSF coefficients with generalized power-law models and found that a softened power law provided the most accurate representation. Using this model, we refitted the PSF describing the diffuse scattered light, shown in blue.

In panel~(a), the softened power-law fit follows the main trend for most of the peaks in the unparameterized fit, consistent with the expectations given the logarithmic scale. In panel~(b),  showing the amount of light that is scattered beyond a given distance, the parameterized and unparameterized fits agree well across most distances. This confirms that the soft–power-law parameterization provides a consistent description. Larger deviations occur only at short scattering distances $\le 3$~pixels, where the unparameterized fit predicts about \SI{8}{\percent} more scattered light. We inspected the parameter bounds of the soft-power-law model and found that it was flexible enough to reproduce the increased scattered light at short distances, but that the fitted solutions systematically preferred less scattering. This suggests that the excess scattering in the unparameterized fit arises from overfitting.

Panel~(c) shows the distribution of total scattered light derived from all $100$ fits in the uncertainty analysis, along with the fit result from the full dataset. The total scattered light obtained from the uncertainty analysis is \SI{41.2 \pm 1.2}{\percent}, compared to \SI{41.1}{\percent} for the fit from the full dataset. Because the fit from the entire dataset incorporates the complete information available, we adopt it as the final PSF solution. Finally, panel~(d) presents this final PSF describing the diffuse scattered-light component.

\subsection{Accuracy and robustness} \label{subsec:diffuse_eval}

\begin{figure}
    \centering
    \includegraphics[width=\textwidth]{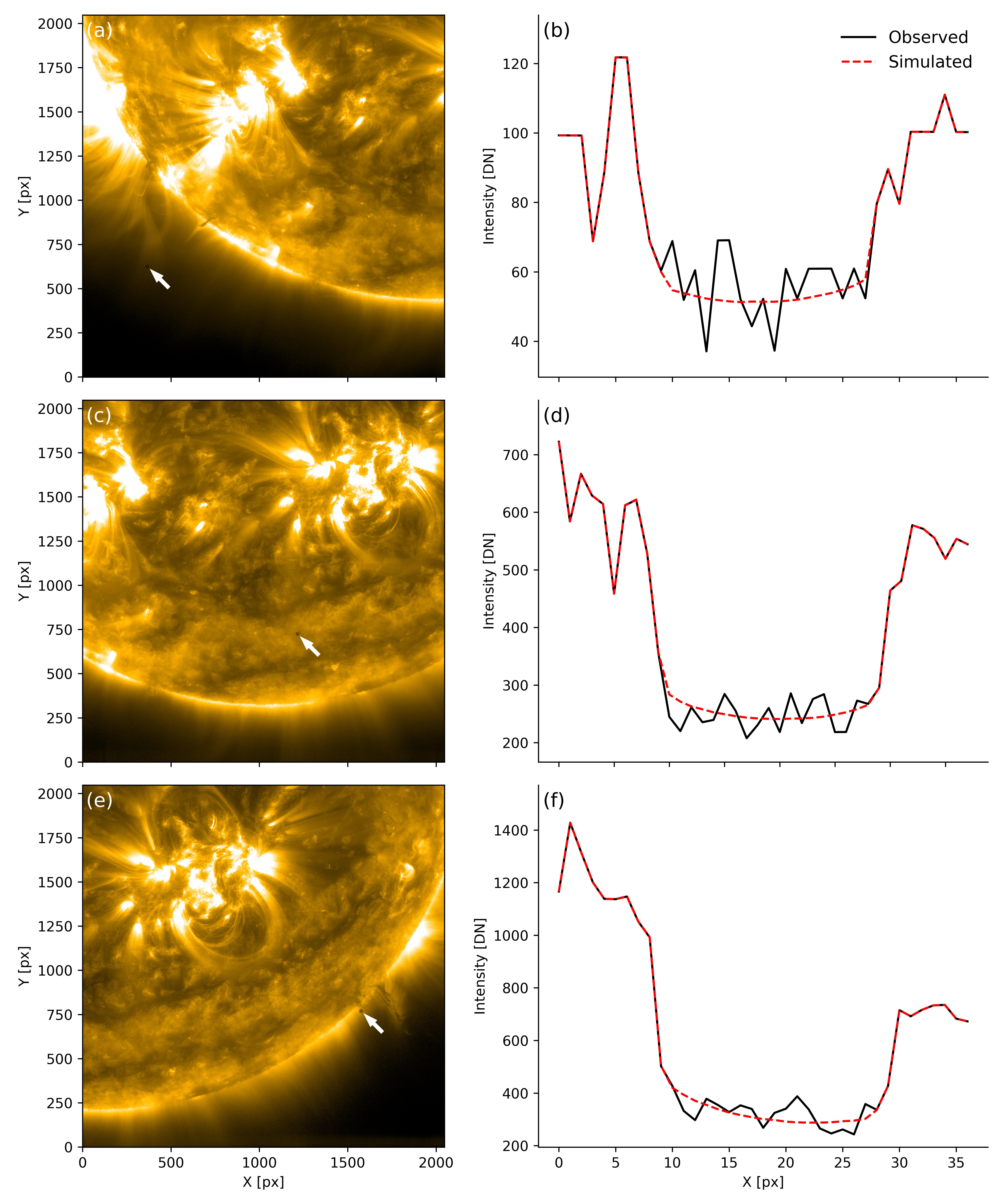}
    \caption{Accuracy of the fitted PSF, evaluated on the Mercury transit occultations observed on (a)--(b) 3~Jan~2023 03:45:27~UT, (c)--(d) 05:57:20~UT, and (e)--(f) 07:32:43~UT. The left column shows the Mercury transit image with Mercury's location marked by a white arrow. The right column shows the observed scattered light and the simulated scattered, derived using the fitted PSF, along horizontal slices through the Mercury occultation.}
    \label{fig:tail_eval_transists}
\end{figure}

\begin{figure}
    \centering
    \includegraphics[width=\textwidth]{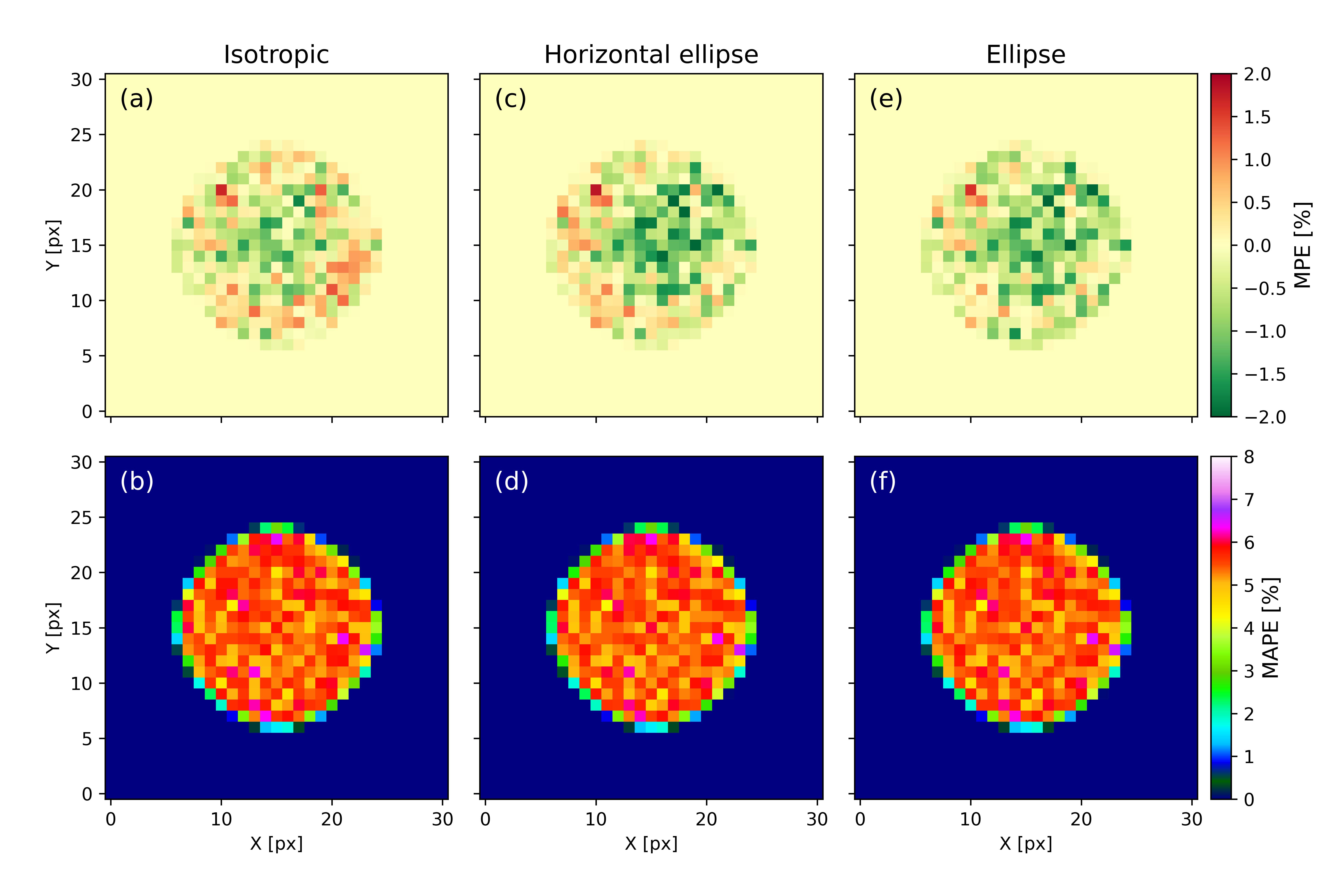}
    \caption{Errors in the simulated scattered light, as derived from a superposed-epoch analysis on all Mercury transit images, assuming (a)--(b) an isotropic PSF, (c)--(d) a PSF parameterized by a horizontal ellipse, and (e)--(f) a PSF parameterized by an ellipse with unconstrained orientation. The first row shows the MPE and the second row the MAPE in the simulated scattered light, dependent on the pixel location within the Mercury occultation.}
    \label{fig:tail_shapes}
\end{figure}

\begin{figure}
    \centering
    \includegraphics[width=\textwidth]{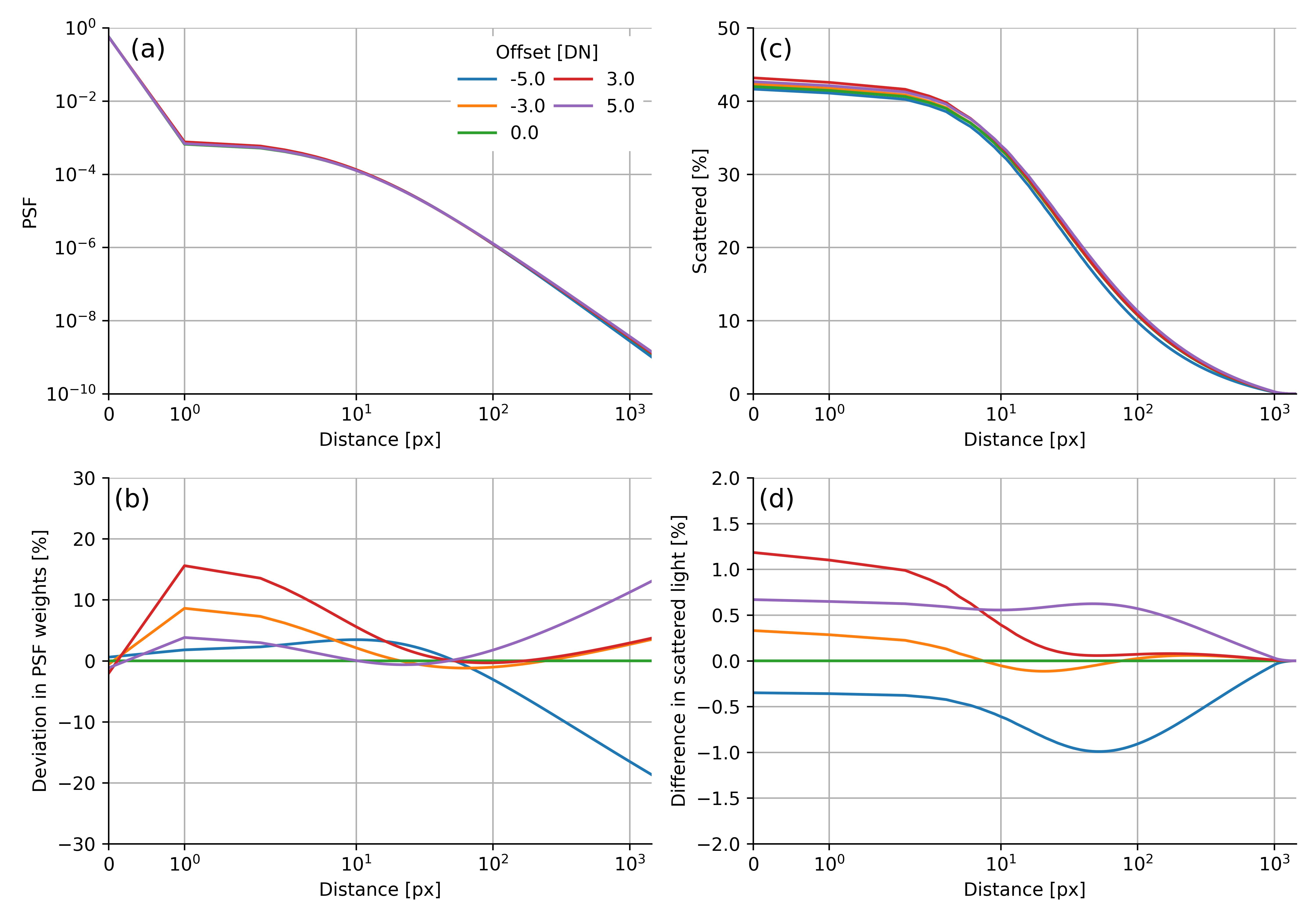}
    \caption{Simulated effect of dark-current calibration offsets on the derived PSF. (a)~Derived PSF weights for assumed calibration offsets of $-5$, $-3$,  $0$, $+3$, and $+5$\,DNs. (b)~Relative deviations of the derived PSF weights from the nominal fit with no calibration errors. (c)~Corresponding amount of light that is scattered farther than a given distance. (d)~Percentage point differences in the amount of light that is scattered beyond a given distance to the nominal fit.}
    \label{fig:tail_pedestals}
\end{figure}

To evaluate the accuracy and robustness of the fitted scattered light component, we performed three experiments: we simulated scattered light into the Mercury occultations and compared it with that observed; we tested for anisotropic scattering; and we tested the effect of uncertainties in the dark current calibration.

To simulate scattered light into the Mercury occultations, we first approximated the true image by deconvolving the observed images with the full fitted PSF, i.e., the merged diffuse and diffraction PSFs, and setting the intensities in the occulted pixels to zero. We then convolved this image with the PSF to simulate the scattered light entering the occultations. Finally, we compared the simulated scattered light to that observed. To ensure a robust evaluation, we applied this procedure in a cross-validation approach: for each of the $100$ PSF solutions from the uncertainty analysis, we simulated the scattered light only for the images in the corresponding evaluation block, i.e., those not used to derive that PSF. The results for three representative cases are given in Figure~\ref{fig:tail_eval_transists}:  when Mercury was far off-limb, on the solar disk, and near the solar limb. In all cases, the simulated scattered light reproduces the observed trends well; the intensity fluctuations in the observations are assumed to arise from photon noise.

We next examined whether the diffuse scattered light shows any significant anisotropy by comparing the spatial distribution of simulated and observed scattered light in the Mercury occultations. For each occulted pixel $\pmb{r}$ in the occulted region (occ), we define the percentage error (PE) and absolute percentage error (APE) as
\begin{align}
    \text{PE}_{\pmb{r}} &= \frac{ I_{\text{sim},\pmb{r}} - I_{\text{o},\pmb{r}}}{\sum_{\pmb{r} \in \text{occ}} I_{o,\pmb{r}}}, \\ 
        \text{APE}_{\pmb{r}} &= \frac{ \left|I_{\text{sim},\pmb{r}} - I_{\text{o},\pmb{r}}\right|}{\sum_{\pmb{r} \in \text{occ}} I_{o,\pmb{r}}}.
\end{align}
We deviate here from the standard convention by normalizing to the average observed intensity over the occulted region. This avoids divergence of the PE and APE when the observed intensities approach zero due to photon noise and is justified since the observed intensity across the Mercury occultation is approximately constant, as can be seen in Figure~\ref{fig:tail_eval_transists}. We then performed a superposed epoch analysis by aligning and stacking all the Mercury occultations. For each pixel position $\pmb{r}$ in the stacked occultation, we computed the mean of the PEs (MPE) and the mean of the APEs (MAPE) from all individual occultations. The results are shown in Figure~\ref{fig:tail_shapes}(a) and~(b). The MPE is typically below \SI{1}{\percent}, indicating that our inversions converged well to the observed solution, and the MAPE is below \SI{6}{\percent}, consistent with the intensity fluctuations attributed to photon noise in Figure~\ref{fig:tail_eval_transists}.
Importantly, the MPE and MAPE show no significant large-scale gradients across the occultation disk, suggesting that the assumption of isotropic diffuse scattering is well justified. 

To further investigate potential anisotropies, we repeated the PSF inversion using spatially stretched power-law parameterizations. First, we considered an ellipse constrained to stretch the PSF only along the horizontal detector axis, consistent with the telescope geometry, with the resulting evaluation shown in Figure~\ref{fig:tail_shapes}(c) and~(d). Second, we considered an ellipse with unconstrained orientation, with the corresponding evaluation shown in Figure~\ref{fig:tail_shapes}(e) and~(f). For each of these two test cases, we subdivided the concentric PSF shells into \SI{15}{\degree}-wide angular segments, fitted the unparameterized PSF over three iterations, and then applied the directionally stretched softened power-law parameterizations for the following three iterations. The resulting inversions showed that elliptical distortions are small, with \SI{0.5}{\percent} horizontal stretching for the constrained case and \SI{0.5}{\percent} stretching at \SI{-15}{\degree} for the unconstrained case. The accuracy of these stretched PSFs, as evaluated by the MPE and MAPE of the simulated scattered light across Mercury's disk analogous to the isotropic case, is illustrated in Figure~\ref{fig:tail_shapes}~(c)--(f). The MPEs are slightly higher and the MAPEs are negligibly lower as compared with the isotropic fit, which is consistent with mild overfitting due to the increased number of free parameters. Overall, the close agreement across all three cases suggests that any anisotropy in the diffuse scattered light is too small to be resolved.

Finally, we evaluated how uncertainties in the dark-current calibration influence the fitted diffuse scattered light. Because the observed intensities in the far off-limb Mercury occultations are only a few tens of counts, systematic calibration errors could bias the results. To quantify this, we repeated the PSF fits assuming systematic dark-current offsets of $-5$, $-3$, $+3$, and $+5$~digital numbers (DN).
Figure~\ref{fig:tail_pedestals} in panel (a) shows the resulting PSF weights for all dark-current offsets. Panel~(b) shows the relative deviations of the derived PSF weights from the nominal fit with no offset. Panel~(c) presents the corresponding amount of light that are scattered beyond a given distance and panel~(d) provides the percentage-point differences in the amount of light scattered beyond a given distance to the nominal fit. Systematic calibration errors change the fitted PSF weights by up to \SI{19}{\percent} relative to the nominal fit, depending on the scattering distance. This, however, only slightly changes how much light is scattered beyond each distance, with differences to the nominal fit remaining below $1.2$~percentage points of the incoming light at all distances. The effect of this $1.2$~percentage-point uncertainty is limited, as it only marginally reduces the sharpness of the reconstructed images and slightly affects the reconstructed intensities in the brightest and darkest image regions.

\section{The combined PSF of \hrieuv} \label{sec:finalPSF}

\begin{figure}
    \centering
    \includegraphics[width=\textwidth]{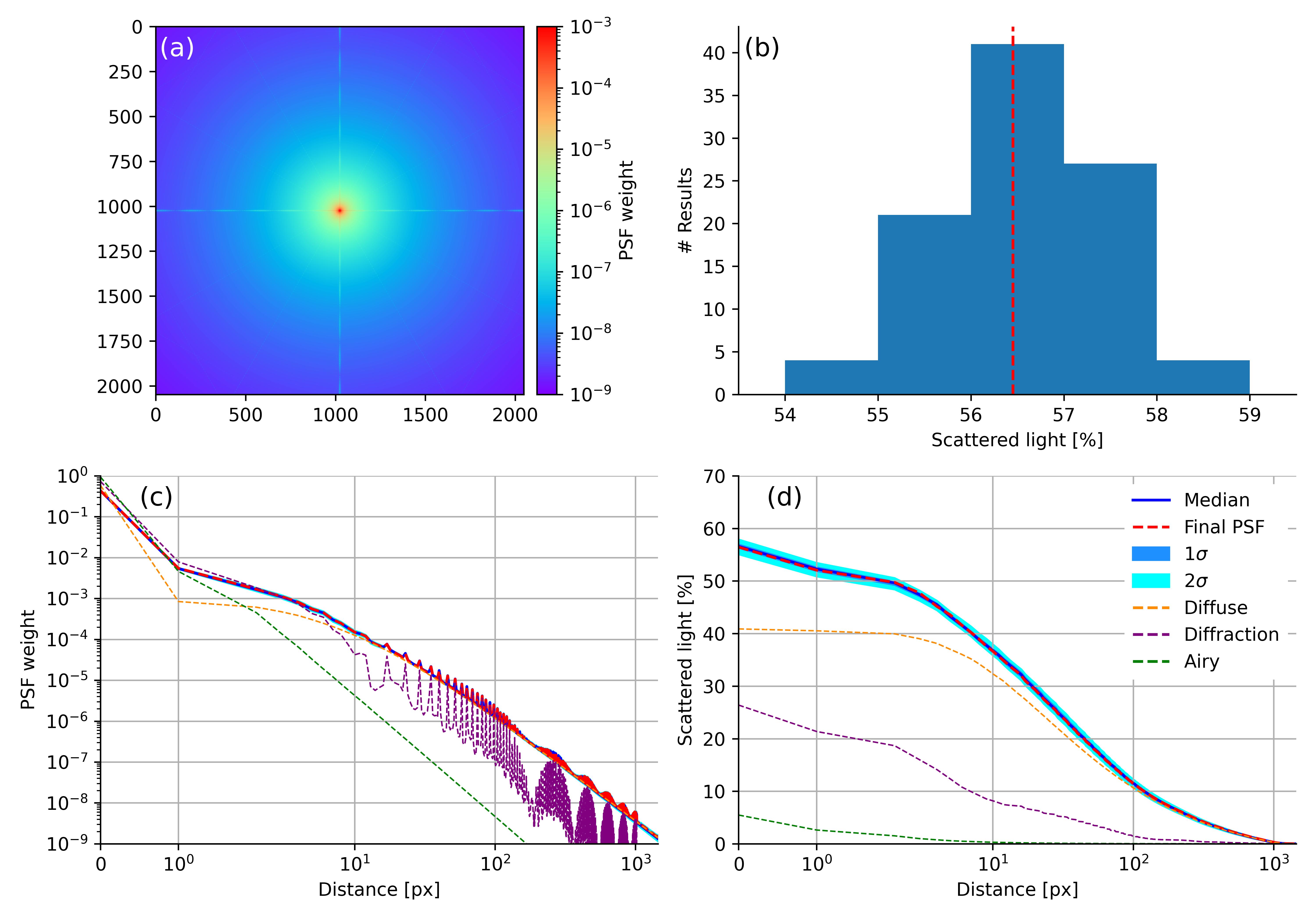}
    \caption{The final \hrieuv\ PSF. (a)~Image of the PSF. (b)~Histogram of the total amount of scattered and diffracted light, as derived from the uncertainty analysis in Section~\ref{sec:diffuse}. The red line indicates the amount of scattered and diffracted light as derived from the entire dataset. (c)~Fitted PSF weights as function of distance from the PSF center, and (d)~amount of light that is scattered and diffracted farther than a given distance. In panels (c)-(d), the red dashed lines shows the fit result as derived from the entire dataset, and the shaded blue curves and bands show the median, $1\,\sigma$, and $2\,\sigma$ confidence levels as derived from the uncertainty analysis. In panels (c)-(d), the yellow curves show the diffuse scattered-light PSF, the purple curves show the diffraction PSF, and the green curves show the ideal diffraction-limited PSF given by the Airy disk. The legend in~(d) is also valid for panel~(c).}
    \label{fig:psf_total}
\end{figure}

The full PSF of \hrieuv\ consists of the diffraction PSF combined with the scattered light PSF. In Figure~\ref{fig:psf_total}, we present this combined PSF including its uncertainties. Panel~(a) shows the PSF, which clearly exhibits diffuse scattering and diffraction from the meshes. The large-scale fine structure from the diffraction due to the spider mount is exceeded by the diffuse scattered light and thus not visible, but the spider mount still contributes significantly to the PSF core. In panel~(b), we show the total amount of diffracted and scattered light as derived from the entire dataset, as well as its distribution  as derived from the uncertainty analysis in Section~\ref{sec:diffuse}. In total, \SI{56.6}{\percent} of light is diffracted and scattered in EUI as derived from the entire dataset, as compared to \SI{56.7 \pm 0.9}{\percent} as derived from the uncertainty analysis. In panel (c), we show the cylindrically averaged PSF weights. The PSF weights range from $4\times10^{-1}$ in the PSF core to $1.5\times10^{-9}$ in the PSF tail, with the diffraction pattern visible. Integrating the PSF weights let us determine the total amount of light that is scattered and diffracted further than a given distance, as is shown in panel~(d). This shows that \SI{37}{\percent} of the incident light is scattered and diffracted farther than \SI{10}{pixels}, \SI{12}{\percent} of the light farther than \SI{100}{pixels}, and \SI{<1}{\percent} farther than \SI{1000}{pixels}. The $1\,\sigma$ and $2\,\sigma$ uncertainties indicate that the PSF is robust, resulting in only minor uncertainties in the image reconstructions.

Figure~\ref{fig:psf_total}(c)–(d) also decomposes the total PSF into its individual components: the diffraction PSF arising from the entrance aperture and  from the entrance and filter-wheel meshes, and the diffuse scattered-light PSF. As a reference, we also show the ideal diffraction-limited PSF, given by the Airy pattern of a circular entrance aperture. In panel~(c), which shows the PSF weights, the diffraction PSF dominates over the diffuse scattered light up to distances of about $7$~pixels; beyond this range, the diffuse scattered-light component becomes dominant. Panel~(d) shows the cumulative fraction of light scattered or diffracted beyond a given distance. About \SI{41.1}{\percent} of the incoming light is diffusely scattered, which is significantly more than the \SI{26.4}{\percent} contribution from diffraction. The convolution of the two components gives the resulting \SI{56.6}{\percent} scattered and diffracted light, as light can be simultaneously diffusely scattered and diffracted. By contrast, the ideal diffraction-limited PSF redistributes only \SI{5}{\percent} of the incoming light, illustrating how significantly diffraction from instrumental structures and micro-roughness scattering shape the observed PSF of \hrieuv.

The PSFs are available online at \url{To Be Done At Acceptance}. To deconvolve \hrieuv\ images with this PSF, we primarily recommend the Richardson-Lucy deconvolution algorithm \citep{richardson1972, lucy1974}. This algorithm is robust for most EUI image-reconstruction use cases and provides stable results, particularly for low-intensity regions. If the target of the analysis is an especially bright region close to the image edge, such as an active region or a flare, we recommend instead the basic iterative deconvolution algorithm \citep{hofmeister2023}, which is more accurate when a large amount of light is scattered beyond the image edge. We note that, to deconvolve an image, one must pad both the image and the PSF with zeros equal to half the image edge length at all sides of the image to break the periodic boundary conditions involved with the convolution in the Fourier space. Python and IDL codes for image deconvolution for various use cases are available at \url{https://github.com/stefanhofmeister/PSF-Tools}. This PSF with the Richardson-Lucy deconvolution can also be invoked directly within the eui\_prep routine for convenience. 

\subsection{Evaluation} \label{sec:evaluation}

\begin{figure}
    \centering
    \includegraphics[width=\textwidth]{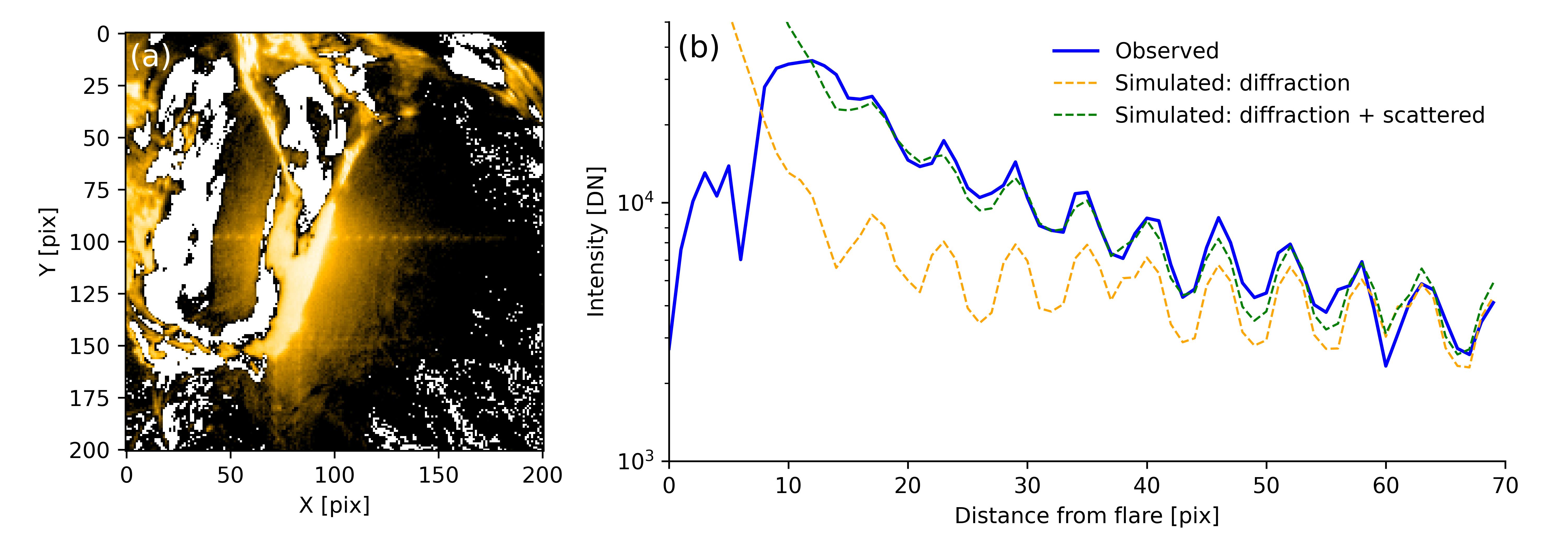}
    \caption{Evaluation on a solar flare. (a)~\hrieuv\ image of an M-class flare on 2024~March~19. The solar background has been subtracted by a \SI{48}{second}-prior non-flaring image. (b)~Observed and simulated intensity distributions along the right diffraction arm near the center of the image. The simulated profiles are derived using the diffraction PSF and the combined diffraction and diffuse scattered light PSF. }
    \label{fig:diff_foreval}
\end{figure}

\begin{figure}
    \centering
    \includegraphics[width=\textwidth]{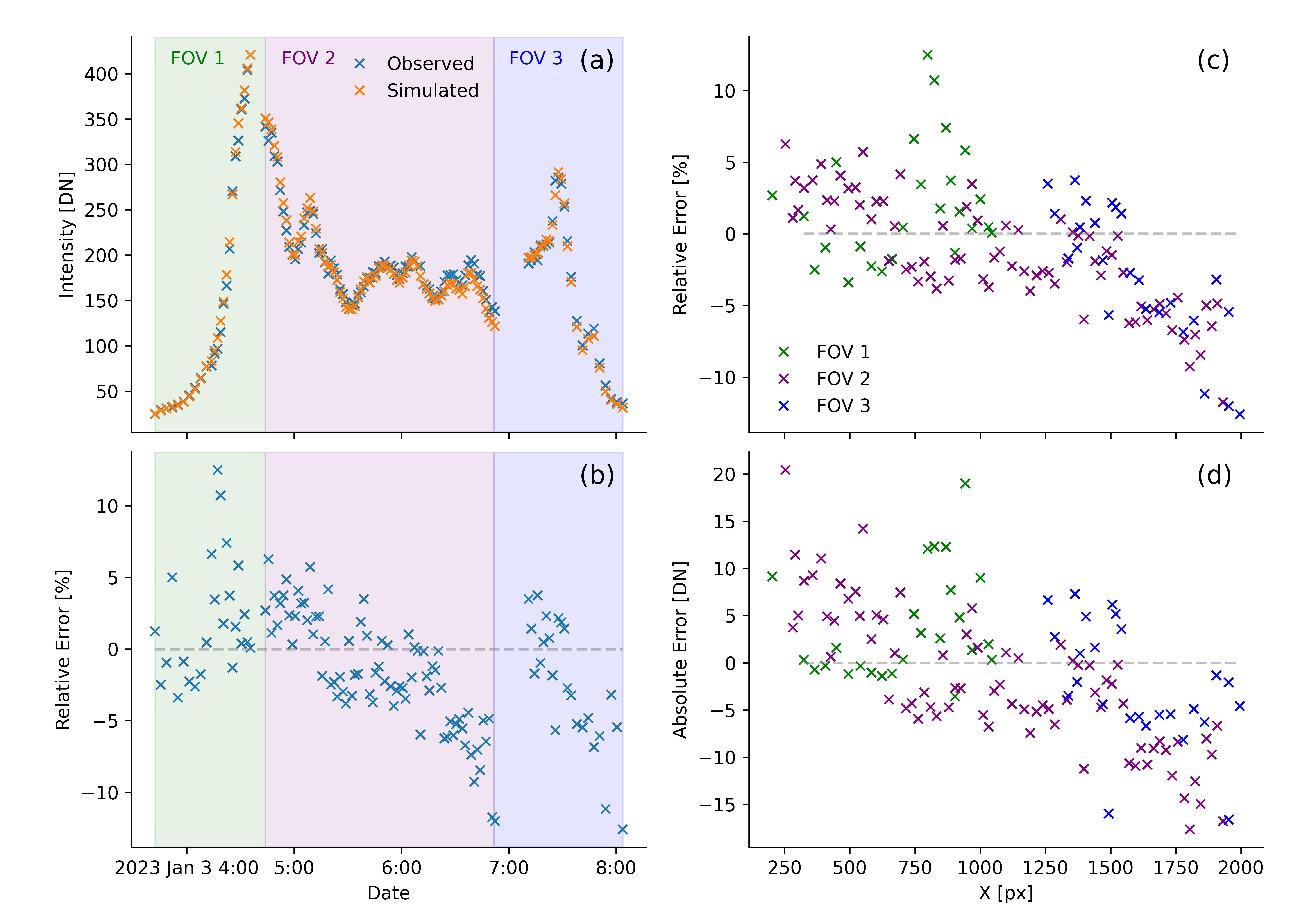}
    \caption{Evaluation using the Mercury transit. (a)~Simulated intensities derived from the diffracted and diffusely scattered light, and the corresponding observed intensities at the center of the Mercury occultation, shown as a function of transit time. (b)~Percentage error in the simulated diffracted and scattered light relative to the observations as a function of transit time. (c)~Percentage error as a function of detector position in the horizontal $X$~direction. (d)~Absolute error as a function of detector position.}
    \label{fig:sim_merc_core}
\end{figure}

\begin{figure}
    \centering
    \includegraphics[width=\textwidth]{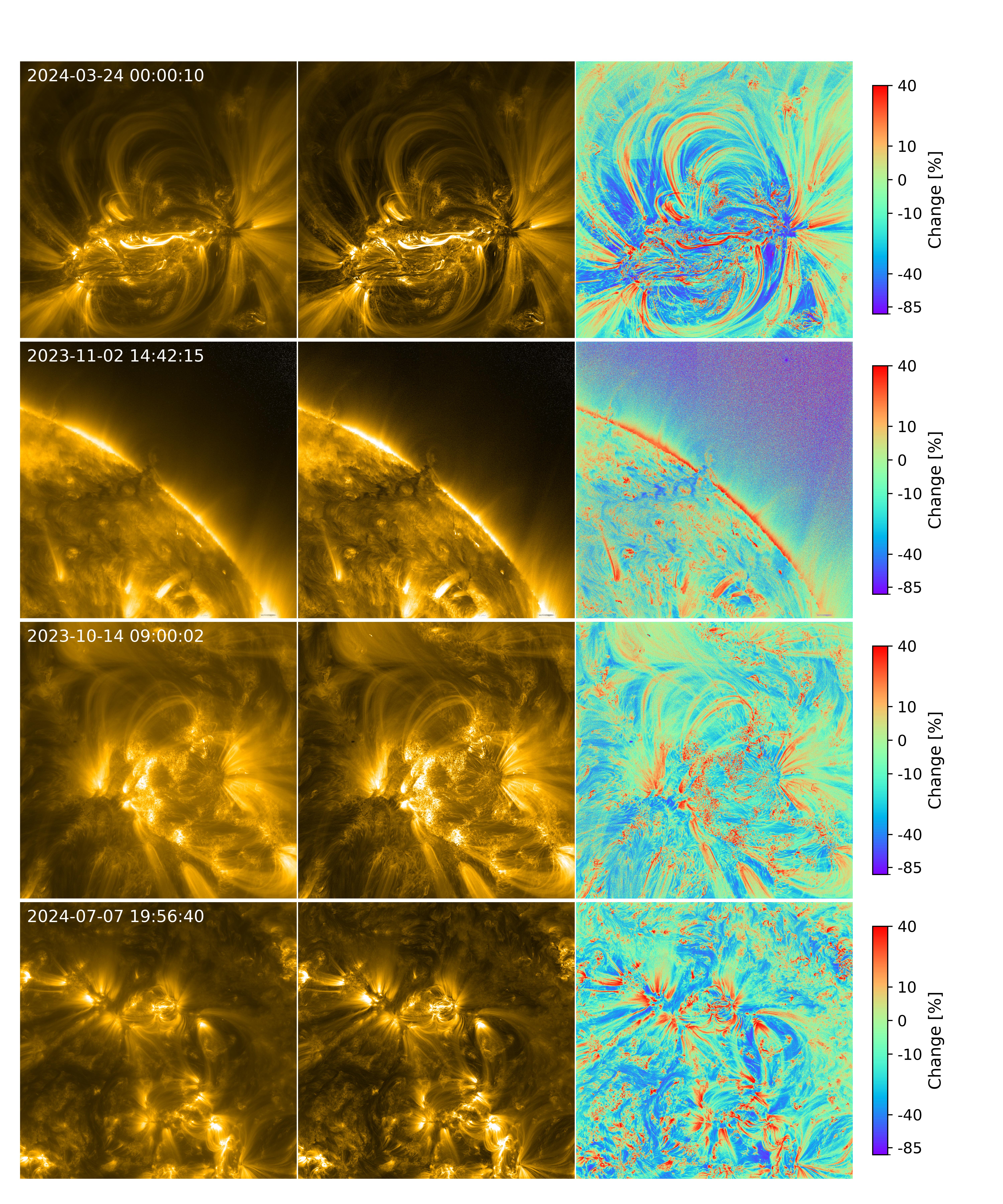}
    \caption{Effect on four representative solar images. The first column shows the original images; the second the PSF-deconvolved images; and the third the percentage change in image intensities.}
    \label{fig:eval_imchange}
\end{figure}
We performed three evaluations: (1) comparing simulated diffracted and scattered light with that observed during solar flares; (2) evaluating the simulated diffracted and scattered light with that observed during the Mercury transit; and (3) assessing how correcting for diffracted and scattered light changes the intensities and dynamic range in the EUV images. 

\subsubsection{Solar flare} 
During solar flares, the diffraction pattern from the meshes and a distinct halo caused by diffuse scattered light become visible around the flare, as shown in Figure~\ref{fig:diff_foreval}(a). These features allow for a basic validation of the PSF by simulating the diffracted and scattered light and comparing it to the observations. The main challenge lies in the saturation of the flare core, which requires simultaneous reconstruction of the flare intensities during the evaluation.

To achieve this, we first isolated the observed diffraction pattern by subtracting the earliest non-flaring image from the flare image. Then, we defined the main flaring location as a $7\times7$~pixel wide box around the center of the isolated diffraction pattern and treated these intensities in the flaring location as free parameters. These intensities were then iteratively adjusted: we simulated the diffracted and diffuse scattered light by convolving the flaring location with the PSF; we compared the simulated diffracted and diffuse scattered light to that observed; and we updated the intensities in the flaring location until the percentage root mean square error was minimized, with the constraint that the mean simulated and observed intensities match for long scattering distances of $60$--$70$~pixels. This procedure was applied to both the diffraction PSF and the final PSF, evaluated along the right diffraction arm in the center of Figure~\ref{fig:diff_foreval}(a). The resulting comparison is shown in panel~(b).

For the diffraction PSF, the simulated diffracted light did not converge well toward the observations, indicating that the diffraction PSF alone cannot reproduce the observed intensity distribution along the diffraction arm. In contrast, the final PSF, which also comprises a diffuse scattered light component, reproduces well the intensity distribution along this diffraction arm, confirming its validity.

\subsubsection{Mercury transit}
The Mercury transit enables us to test whether the scattered light is reproduced equally well at different times and detector positions. This can be tested by simulating the scattered light within the occultations: the images are deconvolved with the PSF, the intensities inside the occulted regions are set to zero to approximate the true images, and the results are reconvolved with the PSF to reproduce the diffracted and scattered light, which is then compared with the observations in the occultation. The result is shown in Figure~\ref{fig:sim_merc_core}. Panel~(a) compares the simulated and observed diffracted and scattered light at the center of the Mercury occultation as a function of transit time. The close agreement, with only minor deviations, indicates that the modeling accuracy does not vary substantially with time or with Mercury’s location.

To examine the deviations in more detail, panel~(b) shows the percentage error of the simulated diffracted and scattered light relative to the observations. The errors are \SI{-1 \pm 4}{\percent} with maximum errors reaching \SI{13}{\percent}, and they vary with time within each EUI targeting field of view. Since time encodes Mercury's location on the detector, this suggests a possible dependence on the detector position. We investigated this further in panels~(c) and~(d), where we present the relative and absolute errors as a function of detector location. The relative errors exhibit a clear positional trend for FOV~2 and~3, albeit with differing slopes, while the absolute errors show a common but more scattered distribution.

A dependence of the relative errors on detector location may indicate that the PSF changes slightly across the detector, whereas a trend in the absolute errors may reflect a small residual intensity gradient. Although the current data do not allow a definitive distinction between these effects, the errors are sufficiently small that they cannot have significantly influenced the PSF determination. Nevertheless, to account for the small spatial variations, the fitted PSF should be regarded as a detector-averaged representation.

\subsubsection{Effects on \hrieuv\ images} 
Finally, we evaluated the impact of the PSF on \hrieuv\ images by deconvolving them and quantifying the percentage change in pixel intensities. This procedure was applied to the \hrieuv\ image catalogue at a daily cadence from May~2020 to December~2024 to visually inspect for artifacts introduced by the PSF deconvolution, but none were found. To illustrate the resulting changes in intensity and dynamic range, we present four examples in Figure~\ref{fig:eval_imchange}. In each case, coronal loops become more sharply defined and the dynamic range increases markedly: bright features intensify by up to \SI{40}{\percent}, while dark regions decrease by up to \SI{85}{\percent}. These results demonstrate that diffuse instrumental scattered light strongly affected the original images, but can be effectively corrected by deconvolution with the PSF.

\section{Discussion } \label{sec:discussion}
We briefly discuss lessons from this study for instrumental design, for observing strategies aimed at characterizing diffuse scattered light, and for potential improvements in follow-up instruments.

\subsection{Diffraction}
Diffraction in \hrieuv\ is primarily determined by the geometry of the spider mount at the entrance aperture and by the meshes supporting the entrance and filter-wheel filters. We first discuss the importance of the locations of the diffracting components in the optical path for PSF derivation, and then address how mesh characteristics affect image formation and correction.

\textbf{Locations of diffracting components}
A fully physical description of diffraction in \hrieuv\ would require Fresnel propagation of the complex wave field from the entrance aperture to the detector. This requirement arises because the filter-wheel mesh is not located at the exit pupil of the optical system. Had the filter wheel been positioned at the exit pupil, both meshes would lie in planes that are Fourier-conjugate to the focal plane, enabling a substantial simplification. In that case, since the exit pupil images the entrance pupil, one could form a combined pupil mask by multiplying the entrance-pupil amplitude mask with the appropriately magnification-scaled exit-pupil mask and then propagate this combined mask to the detector using the Fraunhofer formulation. From a PSF-design perspective, this argues strongly for placing diffraction-relevant components at either the optical entrance or exit pupil of an instrument.

Interestingly, we find that for the filter-wheel mesh, despite the detector being formally located in its near field, the analytical far-field solution provides a good approximation to the resulting diffraction pattern. To assess whether this behavior is specific to the \hrieuv\ geometry, we varied the distance of the filter-wheel mesh from the detector in the one-dimensional telescope modeling described in Section~\ref{sec:diff_implementation}, from its nominal \SI{273}{mm} to $50$, $100$, $200$, $400$, $500$, and \SI{550}{mm}. For all tested distances, the one-dimensional Fresnel near-field solution remained in good agreement with the analytical far-field result. This indicates that, although the diffraction peaks have not yet fully converged in the near field, the angular energy distribution is already well established at these distances. Because the detector pixel size is far too large to resolve individual diffraction peaks, it effectively records the correct integrated signal. We emphasize, however, that this conclusion is instrument dependent, as it depends on the detector pixel size, the mesh geometry, and the distance of the mesh to the detector. While this approximation is likely valid for many EUV imagers used in solar physics to date, it should not be assumed to hold universally for future instrument designs without verification.

\textbf{Mesh geometry}
The mesh geometry provides a degree of freedom to tailor the diffraction characteristics. Increasing the mesh pitch produces a more compact diffraction pattern, redistributing more light over short distances and less over large distances. This slightly concentrates the intensity around the central PSF pixel, which might seem to sharpen the image. However, the enhanced short-range diffraction from the closely spaced peaks  dominates, effectively increasing image blurring and thus reducing sharpness, while the reduced long-range scattering helps preserve the dynamic range. 
In contrast, decreasing the mesh pitch spreads the diffraction peaks farther apart, redistributing light over larger distances. This reduces short-range scattering and thus enhances sharpness, but the stronger long-range scattering degrades more the dynamic range. 

For \hrieuv, a larger mesh pitch was chosen to minimize the total amount of diffracted light. As a consequence, the resulting diffraction peaks are confined to relatively short distances, affecting more sharpness and less dynamic range. However, this confinement also limited our ability to fine-tune the diffraction PSF using observed diffraction patterns from solar flares.

Diffraction patterns can also be exploited to reconstruct saturated flare regions, provided that the diffraction signatures are clearly visible. Because coronal loops in active regions can evolve rapidly during flares, diffraction peaks observed far from the flare, such as in the quiet Sun or, ideally, off-limb, are better suited for reconstructing saturated pixels than peaks located near the active region. A larger diffraction pattern is therefore advantageous, which corresponds to using a smaller mesh pitch. In addition, a smaller mesh pitch increases the separation between diffraction peaks, reducing overlap during spatially extended flares and further improving reconstruction accuracy. Overall, mesh designs with smaller pitches and well-separated diffraction peaks offer the greatest benefit for flare reconstruction.

Larger mesh pitches, however, also offer important advantages. By confining diffraction to shorter distances, they help preserve the dynamic range of the images, which is particularly important when faint structures are the primary science target. In off-limb regions, EUV intensities decrease rapidly, and even a small fraction of light diffracted over long distances from the bright solar disk can dominate the signal from the faint corona. In this context, minimizing long-distance diffraction appears beneficial. However, as shown in this study, the total amount of light diffracted over large distances is substantially smaller than the contribution from diffuse-scattered light, so the impact on dynamic range is secondary. Nevertheless, because the off-limb corona is a key science target for FSI, and because FSI and \hrieuv\ share the same mesh geometry, the need to preserve dynamic range was likely a major factor in selecting a larger mesh pitch for \hrieuv.

Taken together, we argue that a smaller mesh pitch is preferable for solar imagers. For observations of the bright solar disk, a modest increase in long-distance diffraction is acceptable, as it can be accurately modeled from instrument specifications and corrected in post-processing. For off-limb corona observations, well-modeled diffraction generally remains less limiting than the substantial, harder-to-fit diffuse scattered light, making it a secondary concern. At the same time, wider separations between diffraction peaks produce diffraction peaks that are more distinct and detectable over larger distances, facilitating both PSF calibration and the reconstruction of saturated flaring pixels. A smaller mesh pitch reduces the redistribution of light into the central PSF region, enhancing overall image sharpness.

\subsection{Diffuse scattered light}
In EUV imagers, diffuse scattered light is expected to arise primarily from scattering on the microroughness of the telescope mirrors. Although the mirrors are polished to microroughness levels of order \SI{5}{\angstrom}, the short wavelength of EUV radiation (\SI{174}{\angstrom}) is comparable to these surface irregularities, making diffuse scattering relatively efficient.

In principle, diffuse scattered light can be characterized using large-scale occultations, such as lunar transits. In regions deeply occulted by the Moon, any detected signal must originate from very-long-distance scattered light, allowing the tail of the PSF to be well constrained. With these tail coefficients fixed, scattering at progressively shorter distances can then be determined by analyzing intensities closer to the lunar–solar-disk boundary, effectively constraining the PSF from the tail toward the core. This approach requires particular care at very low intensities, where the inferred amount of scattered light depends critically on accurate dark-current calibration and on the linearity of the detector readout at low signal levels.

For EUI, however, no suitable large-scale occultations were available, necessitating the use of the Mercury transit. The small angular size of Mercury introduces a fundamental limitation: it does not allow a clean separation between long-distance scattered light, which defines the PSF tail, and short-distance scattered light, which defines the PSF core. The intensities observed within Mercury's shadow are always a superposition of both contributions. As a result, deriving the diffuse-scattering PSF from a single Mercury occultation is fundamentally ill-posed. Only for the hypothetical case where scattering occurred exclusively on spatial scales smaller than Mercury’s radius would the observed intensity profile in the Mercury occultation uniquely determine the PSF. In the presence of long-distance scattering, however, the inversion admits an infinite number of solutions.

To constrain the solution space, we therefore analyzed a large set of 126 partially occulted images acquired during the Mercury transit. When Mercury is far off-limb, the signal within the occulted region is dominated by long-distance scattered light from the solar disk. When Mercury is on-disk, the occulted intensities are instead dominated by short-distance scattering from the immediately surrounding bright regions. In addition, temporal changes in Mercury’s distance to bright structures, such as active regions, modulate the scattered-light signal in the occulted pixels, further constraining the distance-dependent scattering function. While uniqueness is not guaranteed in principle, our uncertainty analysis based on bootstrapping the input images indicates that this number of observations is sufficient to derive a stable and accurate PSF.

Independent of whether planetary transits or lunar occultations are used, the key requirement for accurately constraining diffuse scattered light is well-calibrated, high-quality signal in the occulted regions. This favors observations with very long exposure times. Saturated regions caused by such exposures can be filled using shorter-exposure images, and partially occulted pixels resulting from the motion of the occulting body during the exposure can be excluded from the analysis. Ultimately, the fidelity of the faint scattered-light signal in the occulted regions primarily determines how accurately the PSF can be recovered, motivating the use of dedicated observing campaigns for PSF characterization, as employed in this study.

\subsection{The \hrieuv\ PSF in comparison to other instruments}

The PSF of an instrument is often regarded as an indicator of overall optical quality. For solar EUV imagers, it is composed of two principal contributions: diffraction and microroughness scattering. Diffraction can be largely controlled through optical and mechanical design and tailored to achieve specific characteristics, making its impact on instrument performance well understood. Microroughness scattering, by contrast, is inherently more complex.

Mirror microroughness is frequently characterized by a single scalar quantity, typically the root-mean-square (RMS) surface height. In reality, however, microroughness is more completely described by the power spectral density (PSD) of surface height variations across spatial scales. This PSD determines the angular scattering function of the diffuse scattered light. Consequently, two mirrors with identical RMS micro-roughness can exhibit substantially different scattering behavior. In addition, the optical design of the instrument plays a critical role. Scattering is fundamentally described by an angular distribution, such that the observed intensity distribution on the detector depends on both the detector size and its distance from the scattering surface. In multi-mirror systems, the scattering introduced by upstream optics is further modified by the focusing or defocusing effects of subsequent mirrors, reshaping the final distribution of scattered light on the detector. As a result, RMS microroughness alone is not a reliable metric for assessing the overall imaging performance of an instrument.

We illustrate this point by comparing \hrieuv\ and the Atmospheric Imaging Assembly (AIA) on board the Solar Dynamics Observatory (SDO). In both systems, the mirror microroughnesses are similar, with a microroughness of about \SI{4}{\angstrom} for AIA and of about \SI{5}{\angstrom} for \hrieuv. Despite this similarity, their scattering characteristics differ substantially. For the AIA \SI{171}{\angstrom} channel, about \SI{16}{\percent} of the light is scattered due to mirror microroughness, while approximately \SI{30.0}{\percent} is diffracted from the meshes (excluding the entrance spider, whose portion could not be resolved for AIA and thus was fitted with the microroughness), resulting in a convolved total of \SI{41}{\percent} scattered and diffracted light \citep{hofmeister2025}. In contrast, \hrieuv\ scatters about \SI{41.1}{\percent} of the light due to micro-roughness and diffracts approximately \SI{26.4}{\percent} (including the entrance spider), yielding a convolved total of \SI{56.6}{\percent}. Given their comparable angular resolutions, these values might suggest that AIA is optically superior to EUI since it scatters and diffracts less incoming light.

However, the spatial distribution of scattered light is equally important. In AIA, the scattered light extends over angular distances of up to \SI{2400}{\arcsec}, whereas in \hrieuv\ it is largely confined within about \SI{500}{\arcsec}. Thus, while \hrieuv\ scatters a larger fraction of the incident light, it does so over much shorter angular distances, improving \hrieuv observations of faint off-limb regions.

This comparison illustrates that the total fraction of scattered light alone is not an adequate metric for evaluating or comparing EUV imagers. Instead, the angular scattering function and its relation to the scientific objectives of the instrument provides a more meaningful basis for comparison. That said, the relatively large fraction of scattered light in \hrieuv\ does indicate a level of microroughness that is high relative to its instrumental design, suggesting that there remains room for improvement in future follow-on instruments.

\section{Conclusion} \label{sec:conclusions}

We have determined the  \hrieuv\ PSF from Mercury transit images. We found that:
\begin{itemize}
\item About \SI{26.4}{\percent} of the incoming light is diffracted by the aperture mesh, the spider mount, and the filter-wheel mesh. The aperture mesh dominates the long-distance diffraction beyond 10~pixels, while all three contributions shape the PSF core at smaller distances.
\item About \SI{41.1}{\percent} of the light is diffusely scattered across the full detector, likely due to mirror micro-roughness.
\item Convolved, these components result in roughly \SI{56.6}{\percent} of the recorded light being diffracted or scattered.
\item Correcting for diffracted and scattered light by deconvolving the collected images with the PSF substantially enhances image sharpness and dynamic range: bright structures intensify by up to \SI{40}{\percent}, and dark structures decrease by up to \SI{85}{\percent}.
\end{itemize}

We have derived the diffraction component from the mechanical drawings, which consists of diffraction from the spider mount and from the entrance-filter and filter-wheel meshes. We found that the spider mount contributes significantly to the PSF core, while the entrance-filter mesh dominates diffraction over large distances. The filter-wheel mesh primarily adds to diffraction over short distances in the PSF core and is negligible over large distances. We could not fine-tune the orientations and spacings of the mesh diffraction patterns due to the lack of sufficiently strong flare images in which several diffraction orders were visible.

In addition, we have determined the diffuse scattered light component from the observed intensities within Mercury occultations. Mercury’s small diameter causes the observed intensity in its center to contain both short- and long-distance diffracted and scattered light from its surroundings, entangling the two components. By simultaneously analysing $126$~transit images, both on-disk and off-limb with varying distances to active regions and to the bright solar limb, we obtained sufficient constraints to fit an initially isotropic scattered-light model, later refined using a softened power-law parameterization.

Lastly, we have tested the robustness of the fit by allowing for anisotropic scattering, by introducing dark-current calibration errors, and by assessing possible PSF variations across the detector. We were unable to determine whether the PSF has a small anisotropic component, a limitation of relying solely on Mercury-transit images with their small occulted regions. Dark-current uncertainties had only a minor effect on the derived PSF, indicating good robustness in that respect. However, the accuracy with which the PSF removes scattered light in the occulted region does depend on Mercury’s position in the image: on average, \SI{99}{\percent} of the observed scattered light is removed with a $1\sigma$ uncertainty of \SI{4}{\percent}, while position-dependent errors can reach up to \SI{13}{\percent}. It remains unclear whether these variations arise from a spatially variant PSF or from a small artificial intensity gradient across the detector, such as a residual light leak. To remain conservative, we therefore recommend interpreting the fitted PSF as a detector-averaged PSF.

 \ \\[.5cm]
We thank the Solar Orbiter EUI team for discussions during the course of this project. Solar Orbiter is a mission of international cooperation between ESA and NASA, operated by ESA. The EUI instrument was built by CSL, IAS, MPS, MSSL/UCL, PMOD/WRC, ROB, LCF/IO with funding from the Belgian Federal Science Policy Office (BELSPO/PRODEX PEA 4000106864,
4000112292 and 4000134088); the Centre National d’Études Spatiales (CNES); the UK Space Agency (UKSA); the Bundesministerium für Wirtschaft und Energie (BMWi) through the Deutsches Zentrum für Luftund Raumfahrt (DLR); and the Swiss Space Office (SSO). 
S.J.H. gratefully acknowledges support from the Solar Orbiter/EUI Guest Investigator program and thanks the EUI team for their kind support during his research stay at the Royal Observatory of Belgium.
S.J.H., M.H. and D.W.S. were supported in part by the National Science Foundation grant AGS-2229100. E.K., C.V. and D.B. acknowledge BELSPO for the provision of financial support in the framework of the PRODEX Programme of the European Space Agency (ESA) under contract numbers 4000143743 and 4000134088.


\facility{Solar Orbiter(EUI)}
\software{Astropy \citep{astropy1,astropy2,astropy3},  
    Matplotlib \citep{matplotlib},
    Numba \citep{numba},  
    Numpy \citep{numpy}, 
    Scikit-image\citep{scikit},
    Scipy \citep{scipy}, 
    Sunpy \citep{sunpy}}

\bibliographystyle{aa} 
\bibliography{bibliography}

\end{document}